\documentclass{emulateapj}

\usepackage{subfigure,epsfig,url,bm,amsmath,cases,color,soul}

\begin{document}

\title{Simulations of the Fomalhaut System Within Its Local Galactic Environment}

\author{Nathan A. Kaib\altaffilmark{1}, Ethan B. White\altaffilmark{1}, \& Andr\'e Izidoro\altaffilmark{2}}

\altaffiltext{1}{HL Dodge Department of Physics \& Astronomy, University of Oklahoma, Norman, OK 73019, USA}
\altaffiltext{2}{UNESP, Univ. Estadual Paulista - Grupo de Din\^amica Orbital \& Planetologia, Guaratinguet\'a, CEP 12516-410 S\~ao Paulo, Brazil}

\begin{abstract}

    Fomalhaut A is among the most well-studied nearby stars and has been discovered to possess a putative planetary object as well as a remarkable eccentric dust belt. This eccentric dust belt has often been interpreted as the dynamical signature of one or more planets that elude direct detection.  However, the system also contains two other stellar companions residing $\sim$$10^5$ AU from Fomalhaut A. We have designed a new symplectic integration algorithm to model the evolution of Fomalhaut A's planetary dust belt in concert with the dynamical evolution of its stellar companions to determine if these companions are likely to have generated the dust belt's morphology. Using our numerical simulations, we find that close encounters between Fomalhaut A and B are expected, with a $\sim$25\% probability that the two stars have passed within at least 400 AU of each other at some point. Although the outcomes of such encounter histories are extremely varied, these close encounters nearly always excite the eccentricity of Fomalhaut A's dust belt and occasionally yield morphologies very similar to the observed belt. With these results, we argue that close encounters with Fomalhaut A's stellar companions should be considered a plausible mechanism to explain its eccentric belt, especially in the absence of detected planets capable of sculpting the belt's morphology. More broadly, we can also conclude from this work that very wide binary stars may often generate asymmetries in the stellar debris disks they host.


\end{abstract}

\section{Introduction}

The Fomalhaut star system has held the interest of planetary astronomers for decades. Beginning in the 1980's, the main star of the system, Fomalhaut A, was found to have an infrared excess, indicating the existence of a significant amount of circumstellar dust \citep{gill86, aum85, backpar93}. Sub-millimeter observations in the next decade suggested that this star's dust disk was actually a belt with a large inner cavity of radius $\sim$100 AU \citep{holl98, dent00}, although newer observations also indicated another less massive dust source 8--15 AU from the star \citep{stap04, su13, su16}. HST observations of the outer dust belt in optical light then revealed that this outer dust belt had a very well-defined inner edge and that the center of the belt was offset from Fomalhaut A, implying that the belt's morphology traced the shape of a moderately eccentric orbit \citep{kal05}. Subsequent observations of the belt in millimeter, sub-millimeter, infrared, and visible light have confirmed that the belt indeed has sharply defined inner and outer edges and has a coherent eccentricity of $0.12\pm0.01$, indicating its source population has a high degree of apsidal alignment \citep{ric12, acke12, mac17, white17}. 

Because of the outer belt's sharp inner edge and non-circular morphology, it has been suspected that Fomalhaut A possesses at least one planet that is gravitationally sculpting the belt's morphology. Secular forcing from a planet orbiting interior to the belt's inner edge can drive the belt to a coherent moderately eccentric shape \citep{wyatt99, kal05, quill06, kal08, chiang09}. Indeed, when \citet{kal08} announced the discovery of a small visible point source orbiting interior to the belt, it was presumed that this was the planet responsible for the belt's morphology. However, failures to detect the object in the infrared raised doubts about the object's mass as well as its planetary nature \citep{marengo09, janson12, currie12}. Further HST observations of the point source at later epochs revealed it to actually be on a highly eccentric orbit incapable of generating the belt's morphology \citep{kal13, beu14}. Moreover, a massive planet on such an orbit would quickly disrupt belt's coherent shape \citep{kal13, beu14, tam14}. Given this, it has since been speculated that it may be a lower mass planet with a large ring system \citep{kal08}, a lower mass planet with an irregular satellite dust cloud \citep{kenwy11, tam14}, a post-collision dust cloud \citep{kal13, tam14, law15}, or even a background neutron star \citep{neu15}. 

Thus, the detected point source near Fomalhaut A does not seem to be responsible for the belt's morphology. Furthermore, it has recently been shown that there is not enough mass in a gaseous component to explain the belt's eccentricity via gas-dust interactions \citep{cat15}. It may be that an undiscovered planet or planets capable of driving the belt's morphology still reside interior to the ring and could even be scattering the detected object onto an extreme orbit \citep{chiang09, tam14}. Another possibility is that a tandem of shepherd planets may reside on each side of the belt driving it into a coherent ellipse \citep{bol12}. Such planets could have masses as low as a few M$_{\earth}$ and easily evade detection \citep{bol12}. 

In addition to the main star, its belt, and the possible planetary body, the Fomalhaut system possesses other known members. As far back as 1938, the K4Ve star TW PsA was suspected of being a binary companion \citep{luyten38}. More recent analysis of the positions, kinematics, and properties of Fomalhaut A and TW PsA find it exceedingly unlikely that the pair are unassociated field stars \citep{mam12}. Instead presumably, the two stars comprise a bound pair separated by $5.74\times 10^4$ AU \citep{barr97, mam12}. Even more recently, a third star, LP 876-10, was identified near Fomalhaut A and TW PsA with similar kinematics, and the probability of it being a field interloper is only $\sim$10$^{-5}$ \citep{mam13}. This third star is an M dwarf sitting $1.58\times10^5$ AU from Fomalhaut A and $2.03\times10^5$ from TW PsA, and it is also presumably a bound member. {\it Herschel} observations of LP 876-10 have revealed that it also possesses a debris disk, a relatively rare occurrence for M dwarfs \citep{kenn14}. For the remainder of this work, we will refer to TW PsA  and LP 876-10 as Fomalhaut B and C, respectively.

Most of the dynamical analysis of the Fomalhaut A system has largely ignored any possible influence exerted by these lower mass stellar companions. The rationale behind this is likely that the stars are inconsequential because their separations are 2--3 orders of magnitude greater than the radius of Fomalhaut A's belt. However, with such large separations, the orbits of these stars are strongly perturbed by the Galactic tide and impulses from other passing field stars \citep{heitre86, hei87, kaib13}. Under these perturbing forces the stellar orbits will cycle through different eccentricities and inclinations. Thus, it is far from assured that Fomalhaut A's stellar companions have never strongly interacted with it. 

One recent work that has not discounted the possible role of Fomalhaut A's stellar companions is that of \citet{shan14}. Seeking to explain the current stellar configuration as well as the eccentricity of Fomalhaut's belt, \citet{shan14} suggested that the Fomalhaut star system originally formed with Fomalhaut A and C comprising a wide binary of $\sim$$10^4$ AU separation, while Fomalhaut B orbited at a much larger ($\sim$$10^5$ AU) separation. Under this scenario, as Fomalhaut B's orbit varies due to the Galactic tide, it usually has close encounters with C, eventually leading to C's ejection while B transitions to a somewhat smaller separation from A ($\sim$$10^{4-5}$ AU) due to conservation of energy. During this process, the belt of Fomalhaut A can be transformed from a roughly circular state to a reasonably eccentric one due to close passages of Fomalhaut A's stellar companions during the reshuffling of stars. 

Although \citet{shan14} demonstrated that Fomalhaut A's stellar companions can excite the eccentricity of its belt, this mechanism relies on a dynamical instability among the stellar orbits that in turn depends on initial conditions that look markedly different from the current stellar configuration. Moreover, this mechanism requires that we are observing the stellar system in a transient state of disruption. Such states are relatively short-lived and after hundreds of Myrs of evolution only $\sim$1\% of these simulated systems should have stellar separations consistent with observations. 

However, it may be possible that Fomalhaut A's companions have not yet passed through an instability. In spite of the system's huge stellar separations, its self-gravity is strong enough to prevent it from being directly stripped apart by the Galaxy's tide \citep{mam13, jiangtre10}. In addition, even through the ratio of stellar separation distances is small, a subset of stellar orbital parameters should allow Fomalhaut B and C to avoid catastrophic scattering events for timescales longer than the system's current age \citep{mam13, holwie99}. Thus, while the current stellar configuration can become unstable at some point, it may have existed in a meta-stable configuration since its birth with a dynamical lifetime longer than 500 Myrs (the system age). 

In such a meta-stable state, it may also still be possible for the stellar companions to strongly perturb the belt of Fomalhaut A. The precession cycling timescale of Fomalhaut B due to C should be at least as long as that due to the Galactic tide \citep{heitre86, beustdutrey06}. As a result, we expect the evolution of Fomalhaut B's orbital eccentricity to be very complex even before considering the random perturbations arising from encounters with other passing field stars \citep[e.g.,][]{rick76}. Under such complex evolution, it is possible that the periastron of Fomalhaut B may have attained very low values at one or more times, allowing Fomalhaut B to pass near enough to the belt of Fomalhaut A to alter its morphology. Thus, even if we take the simplest assumption that the mean stellar separations have always been similar to those observed today, it seems quite possible that Fomalhaut B (or C) could potentially excite the eccentricity of its belt. 

Given the still open question of the mechanism behind Fomalhaut A's eccentric belt, this work will further explore whether its observed state can be an expected outcome of the system's stellar dynamics. Our work is organized into the following sections: In Section 2, we describe the new symplectic routine we have devised to efficiently model planetary dynamics in the presence of 3 or more stellar mass objects. Following this, we describe the details of our simulations modeling the evolution of the Fomalhaut system. In Section 3, we describe the results of our numerical work, detailing the rate that we generate belt morphologies comparable to the observed one and the requirements for doing so. Finally in Section 4, we summarize our work and draw conclusions about the nature of Fomalhaut A's belt and the dynamical history of the system. Also included are two appendices that derive our integration coordinates and quantify the numerical errors of our algorithm.

\section{Numerical Methods}

\subsection{Algorithm Design}

Because of their numerical error-conserving properties while employing large integration step sizes, mixed variable symplectic algorithms are well-suited for numerically modeling the long-term dynamics of planetary systems. However, symplectic modeling of the dynamics of the Fomalhaut system is a distinctly challenging numerical problem. In this system we have belt particles, which follow near-Keplerian orbits in the absence of strong perturbations, as well as three stellar objects. An optimal approach for integrating belt objects and Fomalhaut A is the classical mixed variable symplectic scheme pioneered by \citet{wishol91}. However, complications arise once we consider a planetary system with a stellar mass companion, since the premise of planetary symplectic integrators is that a system is dominated by a single massive body. 

To handle such a scenario, \citet{cham02} devised a symplectic algorithm that can integrate a planetary system along with one distant binary star. In this scheme, planetary bodies (belt objects in our case) are integrated in democratic heliocentric coordinates \citep{dun98} about the primary star, while the wide binary is integrated about the center-of-mass of the system. Using these two reference frames, the binary star is effectively drifted on a Keplerian orbit about the system's center-of-mass, and kicks due to interactions with the planetary system are applied each time step. In such a scheme, the coordinates of the bodies ($\bm{X}$) as a function of the inertial coordinates ($\bm{x}$) are

\begin{equation} \label{eq:chamcoords}
\begin{aligned}
\bm{X}_A &= \frac{m_A\bm{x}_A + m_B\bm{x}_B + \sum_j m_j\bm{x}_j}{m_A + m_B + \sum_j m_j},\\
\bm{X}_i &= \bm{x}_i - \bm{x}_A,\\
\bm{X}_B &= \bm{x}_B - \frac{m_A\bm{x}_A  + \sum_j m_j\bm{x}_j}{m_A + \sum_j m_j}
\end{aligned}
\end{equation}
where the subscripts $A$ and $B$ refer to the primary and secondary stars, respectively, and all other subscripts correspond to planetary objects orbiting the primary. Similarly, the conjugate momenta in this scheme ($\bm{P}$) can be written in terms of the inertial conjugate momenta ($\bm{p}$) as 

\begin{equation} \label{eq:chammom}
\begin{aligned}
\bm{P}_A &= \bm{p}_A + \bm{p}_B + \sum\limits_{j=1}^N \bm{p}_j,\\
\bm{P}_i &= \bm{p}_i - m_i \frac{\bm{p}_A + \sum_j \bm{p}_j}{m_A + \sum_j m_j},\\
\bm{P}_B &= \bm{x}_B - m_B\frac{\bm{p}_A + \bm{p}_B + \sum_j \bm{p}_j}{m_A + m_B + \sum_j m_j}
\end{aligned}
\end{equation}

While the \citet{cham02} algorithm is well-suited for heirarchical systems with two stars, it is still unable to handle systems with 3+ stars. In these star systems, it is less clear what the optimal choice of coordinates for the tertiary star (and beyond) is. Algorithms for planetary integrations within heirarchical systems exist in which the additional stars are integrated in Jacobian coordinates \citep{verev07, beudut04}. However, stellar trajectories diverge further and further from Keplerian as the number of stellar objects increases, and Jacobian integrations are known to break down if close encounters or dynamical reorganization of the system heirarchy occurs \citep{dun98}. Fortunately, implementing a sophisticated coordinate scheme (and Hamiltonian partitioning) may not be necessary for symplectic integrations of the stars. The reason for this is that the simulation time step of a symplectic integration is set by the shortest planetary orbital period in the system. For extended heirarchical stellar systems, the timescale set by the planetary orbits should typically be vastly shorter than the dynamical timescale of the stellar motions. 

While modeling the capture of Oort cloud comets within a star cluster, \citet{lev10} exploited this fact and just integrated cluster stars using a simple $T+V$ leapfrog approach that operated in inertial  coordinates. In \citet{lev10} all objects including comets were integrated with this $T+V$ scheme. While this scheme retained the ability to accurately integrate close encounters (a shortcoming of many symplectic routines), the universal choice of inertial coordinates for all bodies required sacrificing the ability to efficiently integrate planetary bodies assuming near-Keplerian trajectories. This is in contrast to codes such as {\tt Symba} and {\tt mercury}, whose democratic heliocentric coordinates enable rapid planetary integrations and accurate close encounter integrations, but cannot handle more systems with more than two stellar objects \citep{dun98, cham99, cham02}.

However, with a simple modification to Equations \ref{eq:chamcoords} and \ref{eq:chammom} we can use the \citet{lev10} strategy for distant stellar integrations while retaining the advantages of democratic heliocentric coordinates for integrations of bodies in orbit about the primary star. In our new approach, the coordinate system of \citet{cham02} becomes

\begin{equation} \label{eq:mycoords}
\begin{aligned}
\bm{X}_A &= \frac{m_A\bm{x}_A + m_B\bm{x}_B + \sum\limits_{j=1}^{N_P} m_j\bm{x}_j}{m_A + m_B + \sum\limits_{j=1}^{N_P}m_j},\\\\
\bm{X}_i &= \bm{x}_i - \bm{x}_A \quad\quad\quad {\rm for}~1\leq i\leq N_P, \\\\
\bm{X}_B &= \bm{x}_B - \frac{m_A\bm{x}_A +  \sum\limits_{j=1}^{N_P} m_j\bm{x}_j}{m_A + \sum\limits_{j=1}^{N_P} m_j},\\\\
\bm{X}_i &= \bm{x}_i \quad\quad\quad\quad\quad~ {\rm for}~N_P< i\leq N_P + N_S
\end{aligned}
\end{equation}
where $N_P$ is the number of planetary bodies orbiting the primary and $N_S$ is the number of additional stellar bodies beyond the primary and secondary stars. Subscripts from 1 to $N_P$ refer to planetary mass bodies orbiting the primary, while subscripts above $N_P$ refer to stellar mass bodies beyond the primary and secondary stars (which are still represented with subscripts $A$ and $B$ respectively). Thus, we have simply left the additional stars' coordinates in inertial coordinates, while all other bodies use the same coordinates as \citet{cham02}.

Using the generating function detailed in Appendix A, we find that the canonical conjugate momenta for our new coordinates are 

\begin{equation} \label{eq:mymom}
\begin{aligned}
\bm{P}_A &= \bm{p}_A + \bm{p}_B + \sum\limits_{j=1}^{N_P} \bm{p}_j,\\\\
\bm{P}_i &= \bm{p}_i - m_i \frac{\bm{p}_A + \sum\limits_{j=1}^{N_P} \bm{p}_j}{m_A + \sum\limits_{j=1}^{N_P} m_j} {\rm for}~1\leq i\leq N_P,\\\\
\bm{P}_B &= \bm{x}_B - m_B\frac{\bm{p}_A + \bm{p}_B + \sum\limits_{j=1}^{N_P} \bm{p}_j}{m_A + m_B + \sum\limits_{j=1}^{N_P} m_j},\\\\
\bm{P}_i &= \bm{p}_i\quad\quad\quad\quad\quad\quad\quad\quad {\rm for}~N_P< i\leq N_P+N_S
\end{aligned}
\end{equation}

With our new coordinates, we can define two useful positions. The first is the position of the barycenter of primary's planetary system relative to the primary star's position:

\begin{equation} \label{eq:baroffset}
\bm{s} = \frac{\sum\limits_{i=1}^{N_P}m_i \bm{X}_i}{m_A + \sum\limits_{i=1}^{N_P}m_i}
\end{equation}
and the second is the primary star's position in inertial coordinates (expressed in our new coordinates):

\begin{equation}
\bm{\Delta} = \bm{X}_A-\frac{\sum\limits_{i=1}^{N_P}m_i\bm{X}_i + m_B\left(\bm{X}_B + \bm{s}\right)}{m_A + m_B + \sum\limits_{i=1}^{N_P}m_i}
\end{equation}

This small addition to the \citet{cham02} work allows us to split the Hamiltonian into the following terms

\begin{equation} \label{eq:sunpos}
H = H_{\rm Kep} + H_{\rm int} + H_{\rm jump} + T_{\rm S} + V_{\rm S} 
\end{equation}
where

\begin{equation} \label{eq:Hpart}
\begin{aligned}
H_{\rm Kep} = &\frac{P^2_B}{2\mu_{\rm bin}} - \frac{G\mu_{\rm bin}}{R_B}\left(m_A + m_B + \sum\limits_{i=1}^{N_P}m_i\right)\\ 
& + \sum\limits_{i=1}^{N_P} \left(\frac{P^2_i}{2m_i} - \frac{Gm_Am_i}{R_i}\right),\\\\
H_{\rm int} = &-\sum\limits_{i=1}^{N_P}\sum\limits_{j>i}^{N_P}\frac{Gm_im_j}{R_{ij}} + Gm_Bm_A\left(\frac{1}{R_B} - \frac{1}{\left|\bm{X}_B + \bm{s}\right|}\right)\\
& + Gm_B\sum\limits_{i=1}^{N_P} m_i\left(\frac{1}{R_B} - \frac{1}{\left|\bm{X}_B-\bm{X}_i + \bm{s}\right|}\right)\\
& - \sum\limits_{i=1}^{N_P}\sum\limits_{j=N_P+1}^{N_P+N_S}\frac{G m_im_j}{\left|\bm{X}_i - \bm{X}_j + \bm{\Delta}\right|}\\
& - \sum\limits_{i=N_P+1}^{N_P+N_S}\frac{Gm_Bm_i}{\left|\bm{X}_B - \bm{X}_i + \bm{\Delta} + \bm{s}\right|} - \sum\limits_{i=N_P+1}^{N_P+N_S}\frac{Gm_Am_i}{\left|\bm{\bm{X}_i-\Delta}\right|}\\\\
H_{\rm jump} = &\frac{1}{2m_A}\left|\sum\limits_{i=1}^{N_P}\bm{P}_i\right|^2,\\\\
T_{\rm S} = &\sum\limits_{i=N_P+1}^{N_P+N_S}\frac{P_i^2}{2m_i},\\\\
V_{\rm S} = &-\sum\limits_{i=N_P+1}^{N_P+N_S}\sum\limits_{j>i}\frac{Gm_im_j}{R_{ij}},\\
\end{aligned}
\end{equation}

This splitting of the Hamiltonian allows a planetary system with an arbitrary number of stellar companions to be integrated symplectically, with a Wisdom-Holman-like mixed variable symplectic scheme still employed for the secondary star and planetary mass bodies orbiting the primary. For all other stellar bodies, the integration amounts to a simple $T+V$ scheme. The integration kernel for a single timestep, $\tau$, is as follows:

\begin{enumerate}
\item{Advance $H_{\rm int} + V_{\rm S}$ for $\tau/2$}
\item{Advance $H_{\rm jump}$ for $\tau/2$}
\item{Advance $H_{\rm Kep} + T_{\rm S}$ for $\tau$}
\item{Advance $H_{\rm jump}$ for $\tau/2$}
\item{Advance $H_{\rm int} + V_{\rm S}$ for $\tau/2$}
\end{enumerate}
Thus, we have devised a routine that extends the advantages of mixed variable symplectic integrations of planetary motion to systems containing three or more stars.

Because of the appearance of $\Delta$ and $s$ terms (which are both functions of many bodies' coordinates) in Equation \ref{eq:Hpart}, the actual accelerations associated with $H_{\rm int}$ are quite complicated. For instance, the acceleration of planetary body $k$ due to $H_{\rm int}$ is 

\begin{equation} \label{eq:accelpl}
\begin{aligned}
\frac{dv_{u,k}}{dt} = &-\frac{1}{m_k}\frac{\partial H_{\rm int}}{\partial X_k}\\
= &-\sum\limits_{i\neq k}^{N_P} \frac{Gm_i}{R_{ik}^3}\left(X_k - X_i\right)\\
& -\frac{Gm_Am_B}{m_A + \sum\limits_{i=1}^{N_P}m_i}\frac{X_B + s_u}{\left|\bm{X}_B + \bm{s}\right|^3}\\
& -\frac{Gm_B}{m_A + \sum\limits_{i=1}^{N_P}m_i}\sum\limits_{i=1}^{N_P}m_i\frac{X_B - X_i + s_u}{\left|\bm{X}_B -\bm{X}_i + \bm{s}\right|^3}\\
& + Gm_B\frac{X_B-X_k + s_u}{\left|\bm{X}_B -\bm{X}_k + \bm{s}\right|^3}\\
& -\sum\limits_{i=N_P+1}^{N_P+N_S}Gm_i\frac{X_k - X_i + \Delta_u}{\left|\bm{X}_k - \bm{X}_i + \bm{\Delta}\right|^3}\\
& + \frac{1}{m_A + \sum\limits_{i=1}^{N_P}m_i}\sum\limits_{i=1}^{N_P}\sum\limits_{j=N_P+1}^{N_P+N_S}Gm_im_j\frac{X_i - X_j + \Delta_u}{\left|\bm{X}_i - \bm{X}_j + \bm{\Delta}\right|^3}\\
& - \frac{Gm_A}{m_A + \sum\limits_{i=1}^{N_P}m_i}\sum\limits_{i=N_P}^{N_P+N_S}m_i\frac{X_i - \Delta_u}{\left|\bm{X}_i - \bm{\Delta}\right|^3}
\end{aligned}
\end{equation}
Note that since $\frac{\partial \Delta}{\partial X_k} = -\frac{\partial s}{\partial X_k}$ for all $k$, indirect accelerations on the planetary bodies due to interactions between the additional stars and the secondary star are zero. The acceleration of the secondary star due to $H_{\rm int}$ is similarly tedious

\begin{equation} \label{eq:accelbin}
\begin{aligned}
& \frac{dv_{u,B}}{dt} = ~Gm_A\left(\frac{X_B}{\left|\bm{X}_B\right|^3} - \frac{X_B + s_u}{\left|\bm{X}_B + \bm{s}\right|}\right)\\
& + G\sum\limits_{i=1}^{N_P}\left(m_i\frac{X_B}{\left|\bm{X}_B\right|^3} - \frac{X_B - X_i + s_u}{\left|\bm{X}_B - \bm{X}_i + \bm{s}\right|^3}\right)\\
& + \frac{1}{m_A + m_B + \sum\limits_{i=1}^{N_P}m_i} \times\\
& \left(\sum\limits_{i=1}^{N_P}\sum\limits_{j=N_P+1}^{N_P+N_S}Gm_im_j\frac{X_i - X_j + \Delta_u}{\left|\bm{X}_i - \bm{X}_j + \bm{\Delta}\right|^3}\right)\\
& -\frac{m_A + \sum\limits_{i=1}^{N_P}m_i}{m_A + m_B\sum\limits_{i=1}^{N_P}m_i} \sum\limits_{i=N_P+1}^{N_P+N_S}Gm_i\frac{X_B - X_i + \Delta_u + s_u}{\left|\bm{X}_B - \bm{X}_i + \bm{\Delta} + \bm{s}\right|^3}\\
&- \frac{Gm_A}{m_A + m_B + \sum\limits_{i=1}^{N_P}m_i}\sum\limits_{i=N_P}^{N_P+N_S}m_i\frac{X_i - \Delta_u}{\left|\bm{X}_i - \bm{\Delta}\right|^3}
\end{aligned}
\end{equation}
In practice, when we simulate the Fomalhaut system we find it preferable to treat all non-primary stars as additional stars and leave the special coordinate and momentum reserved for the secondary empty (see Appendix B). For reasons of completeness and potential future use we still include the treatment of a ``special'' secondary star in our algorithm description. 

\subsubsection{Galactic Tide}

Because of the enormous separations of Fomalhaut's stellar companions, our code must include perturbations from the Milky Way. To do this, we model the Galactic disk as a slab of density $\rho_0=0.1$ M$_{\sun}$/pc$^3$ \citep{dun87, holmflynn00}. This generates a galactic tide that is dominated by a vertical term relative to the Galactic plane, which leads to additional potential terms in our system:

\begin{equation} \label{eq:galpot}
\begin{aligned}
V_{\rm gal} =& 2\pi G\rho_0 \bigg[m_A\Delta_z^2 + m_B\left(X_{z,B} + s_{z} +\Delta_{z}\right)^2 + \\
& \sum\limits_{i=1}^{N_P} m_i \left(X_{z,i} + \Delta_z\right)^2 + \sum\limits_{i=N_P+1}^{N_P+N_S} m_iX_{z,i}^2 \bigg]
\end{aligned}
\end{equation}
Thus, the potential of the Galaxy's disk produces accelerations on planetary bodies of 

\begin{equation} \label{eq:tidepl}
\begin{aligned}
\frac{dv_{z,k}}{dt} =& - 4\pi G\rho_0 \left(X_{z,k} + \Delta_z\right)\\
&+\frac{4\pi G\rho_0}{m_A + \sum\limits_{i=1}^{N_P}m_i}\left[m_A\Delta_z + \sum\limits_{i=1}^{N_P}m_i\left(X_{z,i} + \Delta_z\right)\right]\\
\end{aligned}
\end{equation}
Similarly, the acceleration on the secondary due to the Galactic potential (if employing the \citet{cham02} secondary coordinates) is 

\begin{equation} \label{eq:tidebin}
\begin{aligned}
\frac{dv_{z,k}}{dt} =& -\frac{4\pi G\rho_0}{m_A + m_B + \sum\limits_{i=1}^{N_P}m_i}\times \\
& \Bigg[\left(m_A + \sum\limits_{i=1}^{N_P}m_i\right)\left(X_{z,B} + s_z + \Delta_z\right)-m_A\Delta_z\\
& - \sum\limits_{i=1}^{N_P} m_i\left(X_{z,i} + \Delta_z\right)   \Bigg]
\end{aligned}
\end{equation}
Meanwhile, the acceleration on any additional stars in the system is just
\begin{equation} \label{eq:tidestars}
\frac{dv_{z,k}}{dt} = -4\pi G\rho_0 X_{z,i}
\end{equation}

\subsubsection{Field Star Passages}

We must also account for perturbations by passing field stars. We do this with the stellar passage prescription of \citet{rick08} where encounter rates of various spectral classes of stars are set by the locally observed density and velocity dispersion of each class. Rather than employing the impulse approximation to model the effects of each stellar passage \citep{rick76}, we directly integrate the passages. Each stellar passage is initiated 4 pc away from the Fomalhaut center-of-mass, and the passing star is followed until its distance from the center-of-mass again exceeds 4 pc, at which point it is removed from the simulation. As with our other stellar objects, passing field stars are integrated with a $T+V$ approach in inertial coordinates. On average, our systems undergo 309 stellar passages within 4 pc every 1 Myr.

\subsubsection{Close Encounters Between Bodies}

One advantage of the democratic heliocentric coordinates used for the integration of planetary bodies is that the close encounters between planetary objects can be accurately integrated \citep{dun98}. In \citet{dun98} this is accomplished by including high frequency force terms in $H_{\rm int}$ that are integrated on very small time steps but disappear at large separations between planetary bodies. On the other hand, \citet{cham02} accurately handles close encounters by taking the terms of $H_{\rm int}$ that involve the encountering bodies $i$ and $j$ and smoothly transferring those terms to $H_{\rm Kep}$ as the distance between bodies $i$ and $j$ decreases. When this transfer begins $H_{\rm Kep}$ ceases to be analytically integrable, and a Bulirsch-Stoer integrator must be employed to integrate $H_{\rm Kep}$ during times of close encounters \citep{stobu80}.

If our multi-star systems are in an unstable configuration, it may be necessary to integrate close encounters between stars as well. In this instance, we adopt the approach of \citet{cham99} and smoothly switch the encountering terms in $V_{\rm S}$ over to $T_{\rm S}$ with the same changeover function as \citet{cham99}. During the encounter, the $T_{\rm S}$ terms involving the encountering stars $i$ and $j$ are no longer analytically integrable, and the integration of these terms is handled by a Bulirsh-Stoer integrator for the duration of the encounter, analogous to \citet{cham99}. 

Of course this encounter strategy requires specifying a critical distance between encountering stars $i$ and $j$ below which the changeover from $H_{\rm int}$ to $H_{\rm Kep}$ begins. For planetary bodies this critical distance is typically a factor of Hill radii. However, this approach is ill-suited to multi-star systems because the large masses and separations of the stars make the Hill radii enormous, and the stellar motion can be quite non-Keplerian. Thus, we instead use the stellar encounter timescale to set the critical encounter distance. Because we are integrating our stars with a $T+V$ approach and $T+V$ orbital integrations require hundreds of steps per orbital period for accurate integration in the planetary orbital regime, we initiate the switchover of encounter terms if the encounter timescale $r_{crit}/v_{enc}$ drops below 500 times the simulation time step. In this setup, the velocities of the stars at infinity are assumed to be small and we take $v_{enc}$ to be the star's escape velocity at $r_{crit}$. For the 5-year time steps we use in this work, this yields an $r_{crit}$ of $\sim$800 AU.

It is important to understand the accuracy limits of our new algorithm, and these limits are mostly set by close encounters between bodies of different classes. At present, our code contains no strategies to handle encounters between objects of different classes (e.g. star-planet encounters). In Appendix B, we characterize the numerical errors that arise from such events, and we demonstrate that in spite of these limitations, our code is still quite suitable for exploring the dynamical evolution of the Fomalhaut system.

\subsection{Fomalhaut Simulations}

\subsubsection{Stars-only Simulations}
To model the evolution of the Fomalhaut system, we initially perform 2000 simulations of just the stellar objects in the system: Fomalhaut A with a mass of 1.92 M$_{\sun}$, Fomalhaut B with a mass of 0.73 M$_{\sun}$, and Fomalhaut C with a mass of 0.18 M$_{\sun}$ \citep{mam13}. The only constraint on the stellar kinematics of the system comes from the 3-D separations between the three stars. At present, stars A and B are separated by $5.74\times10^4$ AU and stars A and C are separated by $1.58\times10^5$ AU. Because of the enormous orbital periods and tiny orbital velocities associated with these separations, the orbital trajectories of the stars are completely unconstrained. Thus, we must resort to statistical arguments to choose our initial conditions for our stars. If one assumes that wide stellar companion orbits are drawn from a distribution that yields an isotropic velocity distribution, Fomalhaut-like stellar orbits will be uniformly distributed in $e^2$ and $\cos{i}$, where $e$ is orbital eccentricity, $i$ is orbital inclination \citep{jiangtre10}. Meanwhile, argument of periastron ($\omega$), longitude of ascending node ($\Omega$), and mean anomaly ($M$) should be uniformly distributed. 

\begin{figure} \centering
\includegraphics[scale=0.43]{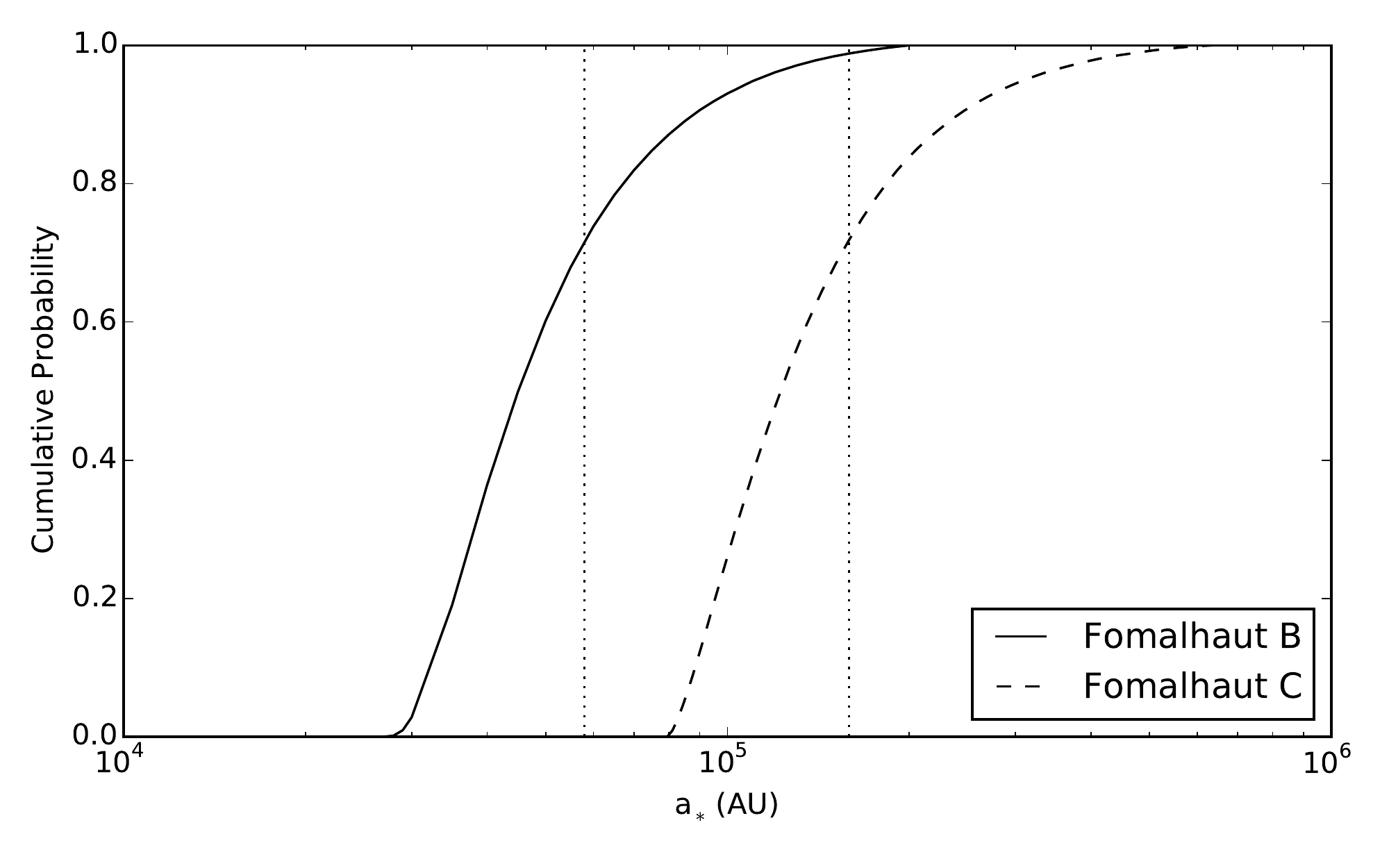}
\caption{Cumulative probability distribution of the semimajor axes of Fomalhaut B ({\it solid}) and Fomalhaut C ({\it dashed}). The nominal 3-D separations of Fomalhaut B from Fomalhaut A and of Fomalhaut C from the system center-of-mass are shown with the left and right vertical dotted lines, respectively.}
\label{fig:SMAdist}
\end{figure}

Using these underlying orbital element distributions, we can then determine the semimajor axis probability function for each star given its current separation. These distributions are shown in Figure \ref{fig:SMAdist}. We see that Fomalhaut B's semimajor axis is likely between $3\times10^4$ to $2\times10^5$ AU (although arbitrarily high values are possible with smaller and smaller likelihood). Meanwhile, the bulk of the probability distribution for Fomalhaut C's semimajor axis lies between $8\times10^4$ to $6\times10^5$ AU. It is these distributions that we sample to assign initial orbital positions and motions to Fomalhaut B and C at the beginning of our simulations. When translating these orbital elements to positions and velocities, we assume that B's orbit describes a trajectory about A, while C's orbit describes a trajectory about the center-of-mass of A and B. Of course, the assumption of orbital elements assumes that each star's trajectory is well-fit by a 2-body solution when in reality the motion can be very non-Keplerian in this system. Nevertheless, we find this approach yields bound systems that continue to match the observed stellar separations for long timescales (see Section 3). With our initial stellar positions and velocities chosen, we next integrate the system in the local galactic environment for 500 Myrs, the approximate maximum age estimate of Fomalhaut A \citep{mam12}. 

\subsubsection{Belt Simulations}
Once our initial 2000 simulations that only include stars are complete, we examine the final positions of the stellar components. If a system finishes with stellar separations within $\pm$50\% of the observed Fomalhaut system (2.9--8.6 $\times10^4$ AU for the AB distance, and 0.79--2.4 $\times10^5$ AU for the AC distance) we consider it a match to the observed system. For systems that match the observed stellar separations, their initial conditions are then reintegrated for 500 Myrs with a belt of 500 massless test particles on initially nearly circular ($e<10^{-3}$), coplanar ($i<1^{\circ}$) orbits between 127 and 143 AU from Fomalhaut A. Because we may be required to integrate the ring to high eccentricity, we use a simulation time step of 5 years, which is just $\sim$1/200 of the smallest ring particle orbital periods.  This has the added benefit of limiting the number of encounters between stellar companions and the primary or ring particles that degrade the accuracy of our simulations (see Figures \ref{fig:secondaryerr}--\ref{fig:starplELerr} in Appendix B). 

\section{Results and Discussion}

\subsection{Stability of Stellar System}

Using stability studies of small bodies within binary systems by \citet{holwie99}, \citet{mam13} argue that there are many combinations of orbital eccentricities for Fomalhaut B and C that enable the system to remain stable. However, as the Fomalhaut system evolves, it is inevitable that its stellar orbits are continuously altered by perturbations from the Galactic tide and passing field stars \citep{heitre86}. Thus, even if the system's stars are born on a stable orbital configuration, they may quickly evolve to an unstable configuration \citep{kaib13}. 

As an alternate origin hypothesis, \citet{shan14} propose that the system originally formed in a tighter configuration, with Fomalhaut C orbiting inside of the A-B orbit. Due to star B's orbital variation under the Galactic tide, the system eventually destabilized, and the ejection of C over the course of tens of Myrs was initiated. \citet{shan14} argue that we are observing the system in the process of C's ejection. By studying the evolution of 1000 systems that start with a compact configuration, \citet{shan14} find that $\sim$1\% of systems will have the observed stellar separations after 400--500 Myrs. 

\begin{figure} 
\centering
\includegraphics[scale=0.43]{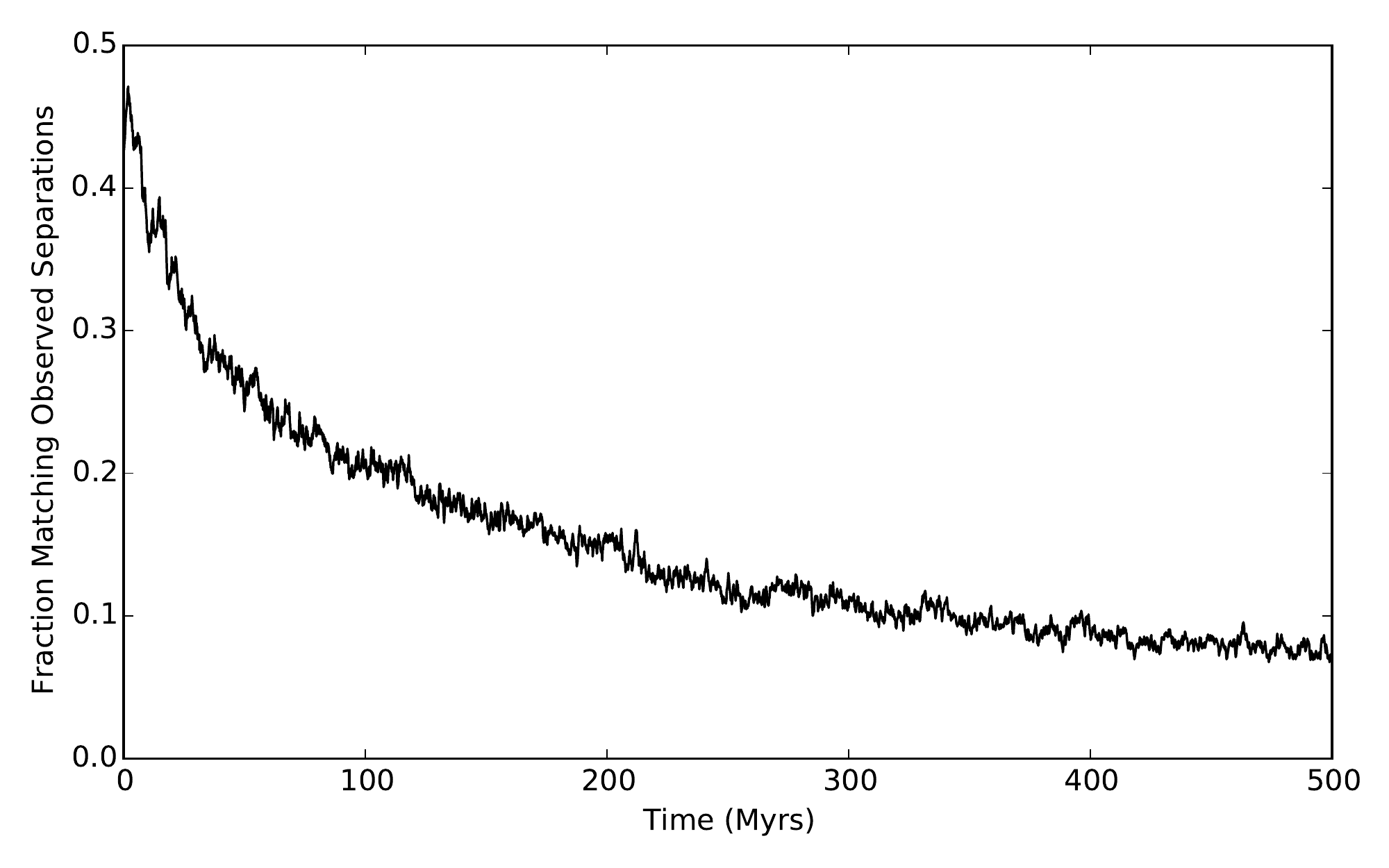}
\caption{Fraction of our 2000 systems with stellar separations that are within 50\% of the 3-D separations reported in \citet{mam13} as a function of time. } \label{fig:matchseps}
\end{figure}

While the stellar kinematics information necessary to test the scenario of \citet{shan14} is well beyond our current observational abilities, we test a simpler scenario here: that the companion stars have always had separations and orbits roughly consistent with the system's current state. To test this, we simply start our stars with the current separations (as described in our numerical setup) and see how many systems retain the approximate separations over the system's age. We do this by plotting the instantaneous fraction of our 2000 systems that have AB separations between $\pm$50\% of $5.74\times10^4$ AU as well as AC separations between $\pm$50\% of $1.58\times10^5$ AU as a function of time. This is shown in Figure \ref{fig:matchseps}. Here we see that our initial conditions cause 43.4\% of our systems to begin within 50\% of the observed separations. (The reason this is not 100\% is that our initial stellar orbits have non-zero eccentricities, and we randomly select the initial mean anomalies of the stars.) As our 2000 systems evolve, the fraction matching the observed separations steadily decreases. By independently selecting stellar semimajor axes and eccentricity, it is inevitable that some of our initial conditions are inherently unstable, but also importantly, the cycling of stellar orbital elements by galactic perturbations drives initially stable configurations into instability. Typically, these instabilities lead to the ejection of star C and a decrease in the mean AB separation due to energy conservation. Nevertheless, after 500 Myrs of evolution 6.75\% of our systems finish with stellar separations that are within 50\% of the observed separations. Moreover, because of the cycling of mean anomaly on eccentric orbits, we expect that many systems regularly oscillate between matching and not matching the observed separations. Indeed, $\sim$51\% of our systems possessed stellar separations within 50\% of the observed values {\it at some point} during their final 100 Myrs of evolution. These fractions are significantly higher than those found in \citet{shan14}, and we therefore consider it quite plausible that the Fomalhaut system has always possessed stellar orbits comparable to those implied by its current stellar separations. 

\subsection{Interactions Between Stellar Companions}

Even in the absence of external perturbations, the large separation of Fomalhaut A and B causes Fomalhaut C's orbital motion to be quite non-Keplerian. On the other hand, because Fomalhaut C is substantially less massive than A and B, their trajectories should be much closer to Keplerian orbits in the absence of perturbations. In reality of course, the Fomalhaut system is continually subjected to velocity impulses from passing field stars and torques from the tide of the Milky Way's disk. Consequently, the motion of both Fomalhaut B and C (relative to A) is distinctly non-Keplerian in our simulations. One manifestation of this behavior is that the distance of closest approach between the stellar companions continually changes as the stars orbit about one another. 

\begin{figure} 
\centering
\includegraphics[scale=0.43]{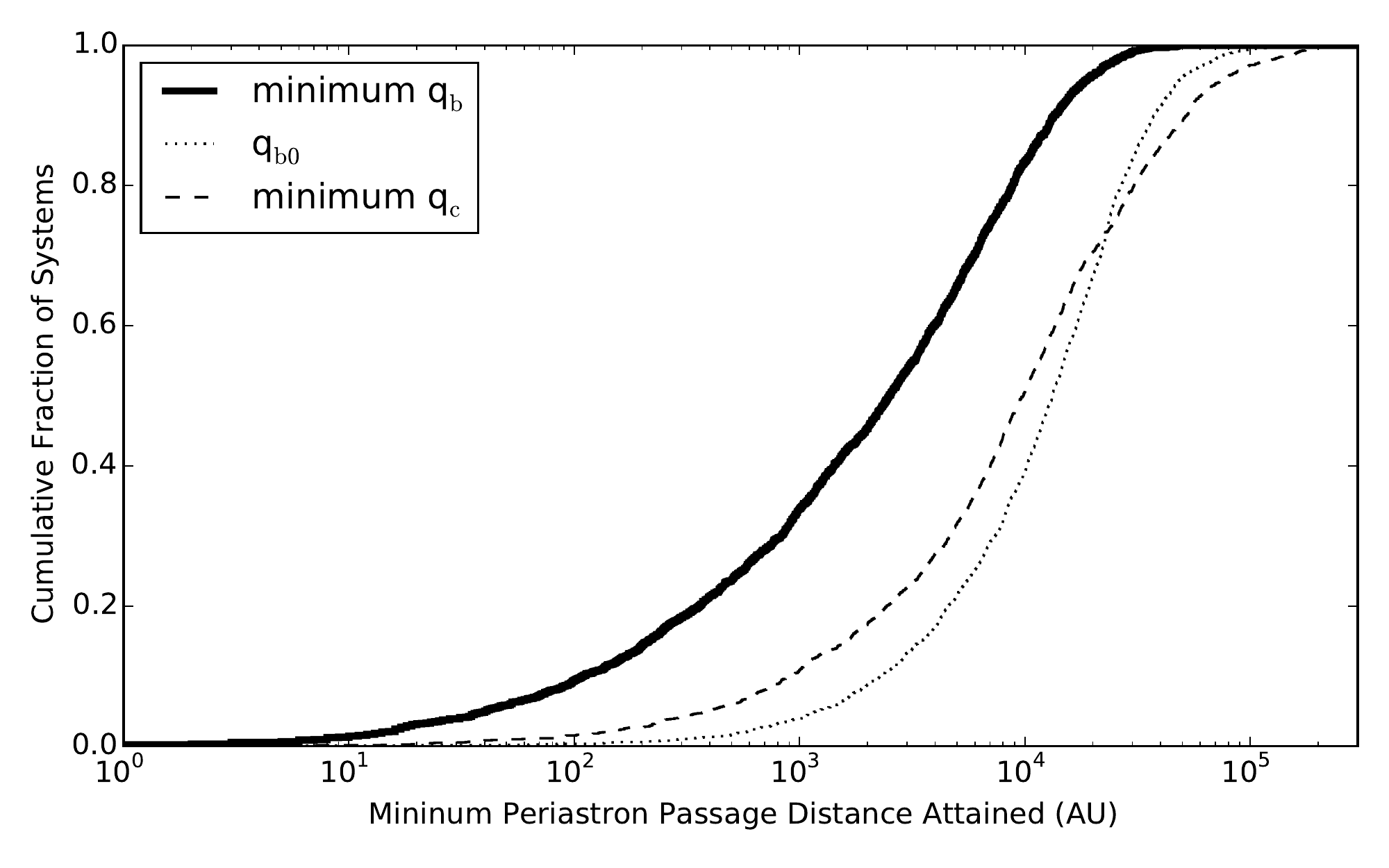}
\caption{The cumulative distribution of the minimum periastron passage for Fomalhaut B attained in each of our 2000 simulated systems is shown with the solid black line. The minimum periastron values for Fomalhaut C are also shown ({\it dotted}) as well as the initial periastron values of Fomalhaut B ({\it dashed}).}
\label{fig:peridists}
\end{figure}

In Figure \ref{fig:peridists} we plot the distribution of the minimum approach distance recorded between Fomalhaut A and B in each of our 2000 simulations. We see here that over the course of 500 Myrs, a typical system experiences at least one approach between A and B that is much closer than that given by the Keplerian orbit initially assigned to B at the beginning of our simulations. In fact the median minimum approach distance between A and B for our simulation set is 2500 AU, and 24\% of simulated systems recorded at least one encounter between A and B below 400 AU. These numbers are substantially more extreme than that specified by our initial conditions. The initial periastra assigned to Fomalhaut B have a median value of $1.32\times10^4$ AU, and only 1.6\% of initial periastra are under 400 AU. Furthermore, the median closest approach between Fomalhaut A and C is 9800 AU, which is a factor of $\sim$16 smaller than their current observed separation. 

We can conclude from this that the complex dynamics of the Fomalhaut system very often yield close approaches between stellar companions, regardless of the system's initial state. If we examine Figure \ref{fig:starpasserr}A in the code validation tests of Appendix B, we see that single passages of Fomalhaut B that are below $\sim$400 AU often excite Fomalhaut A's belt to eccentricities comparable to or greater than the observed value ($\sim$0.1). Thus, this simulation set supports the idea that Fomalhaut A's eccentric belt may be an effect of Fomalhaut B's perturbations. 

\subsection{Belt Morphologies}

\subsubsection{Observed Morphology}

Using new ALMA observations of the belt around Fomalhaut A combined with the expected dust particle locations for a given underlying orbital distribution, \citet{mac17} build a best-fit model for the distribution of orbits found in the belt. Working under the assumption that the belt is shaped by a planetary perturber, the orbital eccentricities and arguments of pericenter are composed of a forced component (associated with the forcing from the planet) and a free component \citep[e.g.][]{murder99}. In the \citet{mac17} model, the forced components all have a singular value for the eccentricity and argument of pericenter, and these components account for the coherent elliptical shape of the belt. Meanwhile, the free components are also given a fixed eccentricity value but have randomly distributed arguments of pericenter. The free components effectively result in a scatter in both the true orbital eccentricities as well as the degree of apsidal alignment when they are added to the forced component. If the free eccentricity is much smaller than the forced, then the ring will be perfectly apsidally aligned with very sharp edges. On the other hand, if the free eccentricity is much larger, the belt's boundaries will be less distinct and its shape will average to a circle. \citet{mac17} find that the best fit model to ALMA data requires a forced eccentricity of $0.12\pm0.01$ and a smaller free eccentricity of $0.06\pm0.04$. 

The modeling of a forced and free eccentricity is tied to the idea that the ring's morphology is shaped by constant perturbations from a planet on a fixed orbit. Meanwhile, we expect the belts in our simulations to be altered by impulse-like periastron passages of stellar companions on highly variable near-parabolic orbits. Thus, the forced/free idealization is not very applicable to our scenario. Nonetheless, this best-fit model of \citet{mac17} predicts the spread of eccentricities and arguments of pericenter (as well as longitudes of pericenter if the bodies are nearly coplanar) to be expected in the observed disk, and these are properties we wish to study and compare in our simulations.

\begin{figure} 
\centering
\includegraphics[scale=0.43]{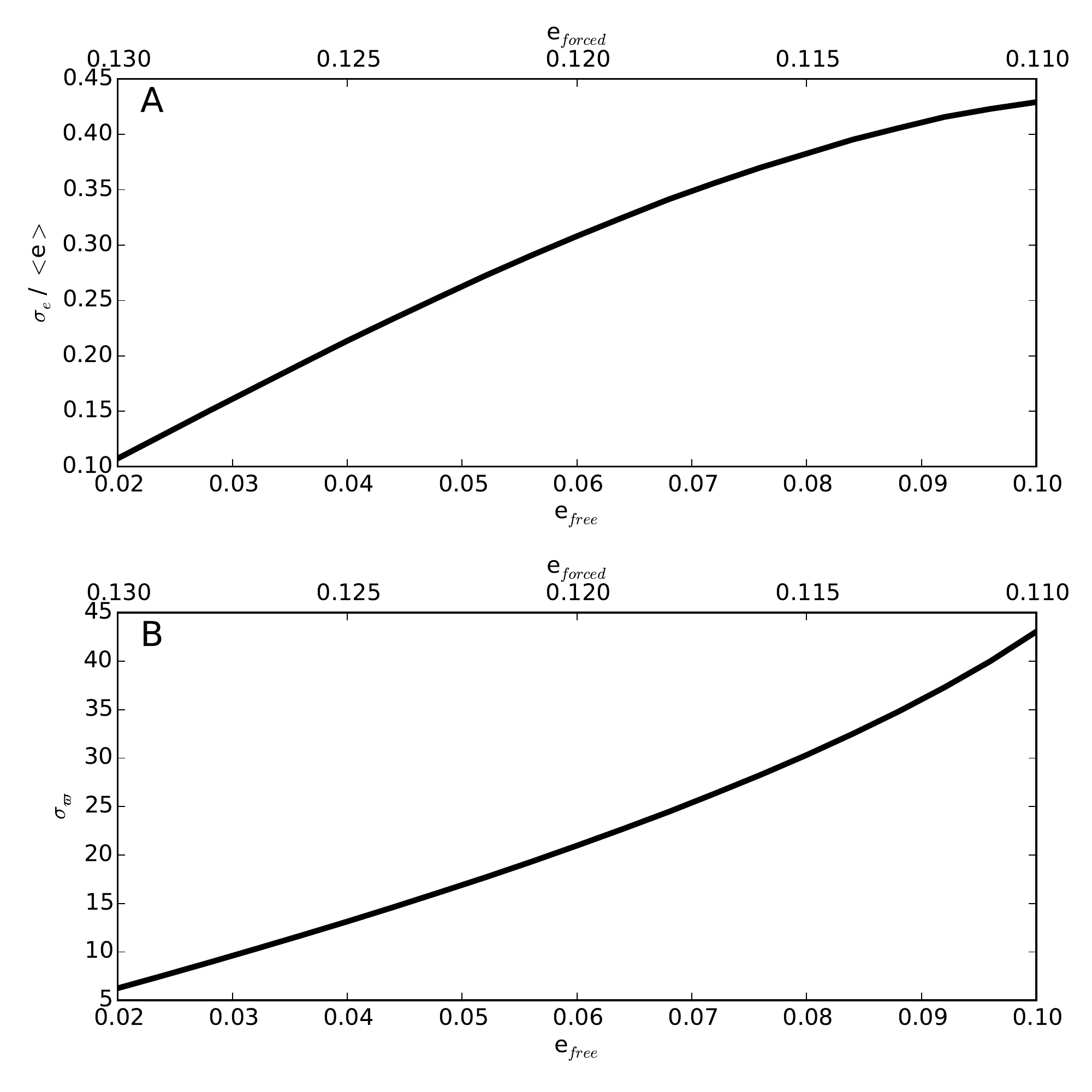}
\caption{{\bf A:} The ratio of the standard deviation of the belt eccentricities to the median belt eccentricity as a function of the observed belt's free/forced eccentricity specified along the $x$-axes according to the best fit belt model of \citet{mac17}. {\bf B:} The standard deviation of the longitudes of pericenter as a function of the observed belt's free/forced eccentricity specified along the $x$-axes according to the best fit belt model of \citet{mac17}.}
\label{fig:modelrange}
\end{figure}

To estimate the best-fit model's expected spread in eccentricity and longitude of pericenter, we calculate these values while varying the forced eccentricity and free eccentricity across the uncertainty of the best-fit model of \citet{mac17}. In Figure \ref{fig:modelrange}A, we plot the expected ratio of the standard deviation of eccentricity to the median eccentricity against different assumed combinations of free and forced eccentricity predicted by the best-fit model. This ratio will be lowest when the free eccentricity is at a minimum and the forced is at a maximum, and it will be highest when the free eccentricity is at a maximum and forced is at a minimum. From this plot we see that the best-fit belt model of \citet{mac17} predicts a belt with a ratio of standard deviation in eccentricity to median eccentricity between 0.11 and 0.43

We can also study how different combinations of free and forced eccentricity will affect the degree of apsidal alignment seen in the best-fit model of the belt. In Figure \ref{fig:modelrange}B, we plot the standard deviation in the longitude of pericenter as a function of the combination of free and forced eccentricity employed in the best-fit model of \citet{mac17}. (We also assume a fixed longitude of ascending node, which would be the case for a near coplanar belt). As can be seen, when forced eccentricity is maximized and free eccentricity is minimized there is a standard deviation of just 6.2$^{\circ}$, indicating apsidal alignment is strongest. The situation is reversed when forced eccentricity is minimized and free is maximized. In this case, the standard deviation is 43$^{\circ}$, indicating weaker alignment. Thus, belts in which the standard deviation of the orbital longitude of ascending node is between 6 and 43$^{\circ}$ provide an acceptable match to the best-fit model of the observed belt of Fomalhaut A.  

\subsubsection{Simulated Morphologies}

Our initial set of 2000 simulations only contain the Fomalhaut stellar companions and do not contain any belt particles. To simultaneously evolve a configuration of belt particles significantly adds to the simulations' computational expense, so we choose to do this only for simulations that yield stellar configurations consistent with the observed present-day system. That is, after our initial 2000 simulations are run for 500 Myrs, we select any systems where the final separation between Fomalhaut A and B is $\pm50$\% of $5.74\times10^4$ AU and the final separation of Fomalhaut A and C is $\pm50$\% of $1.58\times10^5$ AU. These simulations are then rerun for 500 Myrs with a disk of nearly circular, coplanar massless test particles around Fomalhaut A. In total we repeated 135 (6.8\%) of our simulations.

\begin{figure} 
\centering
\includegraphics[scale=0.43]{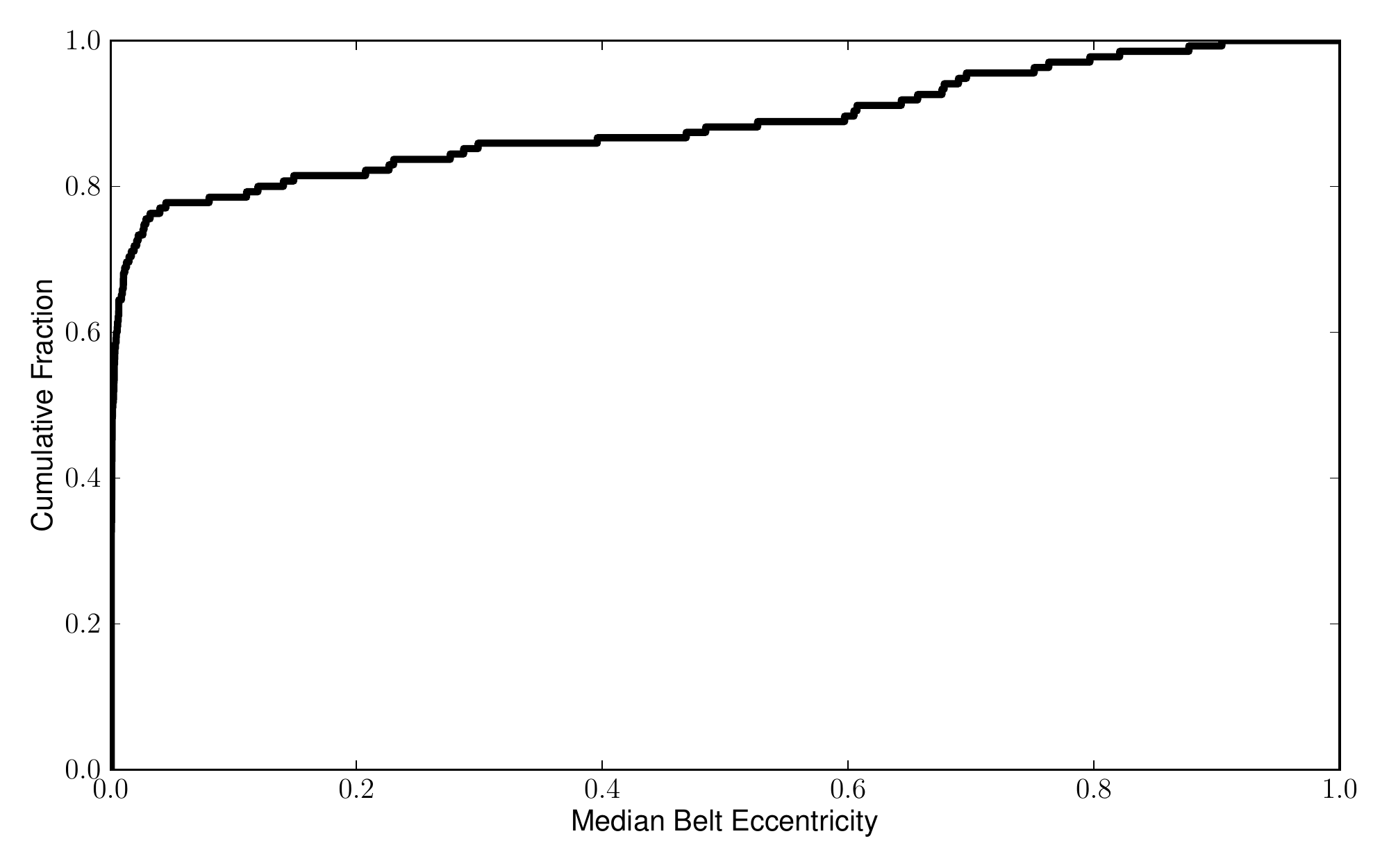}
\caption{Cumulative distribution of the median eccentricity of each of our 135 Fomalhaut A belts after 500 Myrs of evolution.}
\label{fig:cumulecc}
\end{figure}

At the end of these simulations, we examine the orbital distributions seen in our 135 different Fomalhaut A belts. In Figure \ref{fig:cumulecc} we plot the cumulative distribution of the 135 median belt eccentricities that we find. We see a distinct ``knee'' in this distribution around a median eccentricity of 0.04. Roughly 75\% of our simulated belts have median eccentricities between 0 and 0.04, and the remaining 25\% have eccentricities between 0.04 and 1. 

Such a distribution makes sense when we again consider Figures \ref{fig:peridists} and \ref{fig:starpasserr} as well as the fact that $\sim$25\% of our systems experience at least one encounter between Fomalhaut A and B below 400 AU. Thus, this remaining quarter of simulations represent the systems whose belts are significantly perturbed by one or more periastron passages of Fomalhaut B. Among these systems, the distribution of eccentricities between 0.04 and 1 is nearly flat. Thus, system histories that involve close encounters between Fomalhaut A and B are just as likely to yield a belt with an eccentricity of 0.1 as they are to yield a belt eccentricity of 0.8. 

\begin{figure} 
\centering
\includegraphics[scale=0.43]{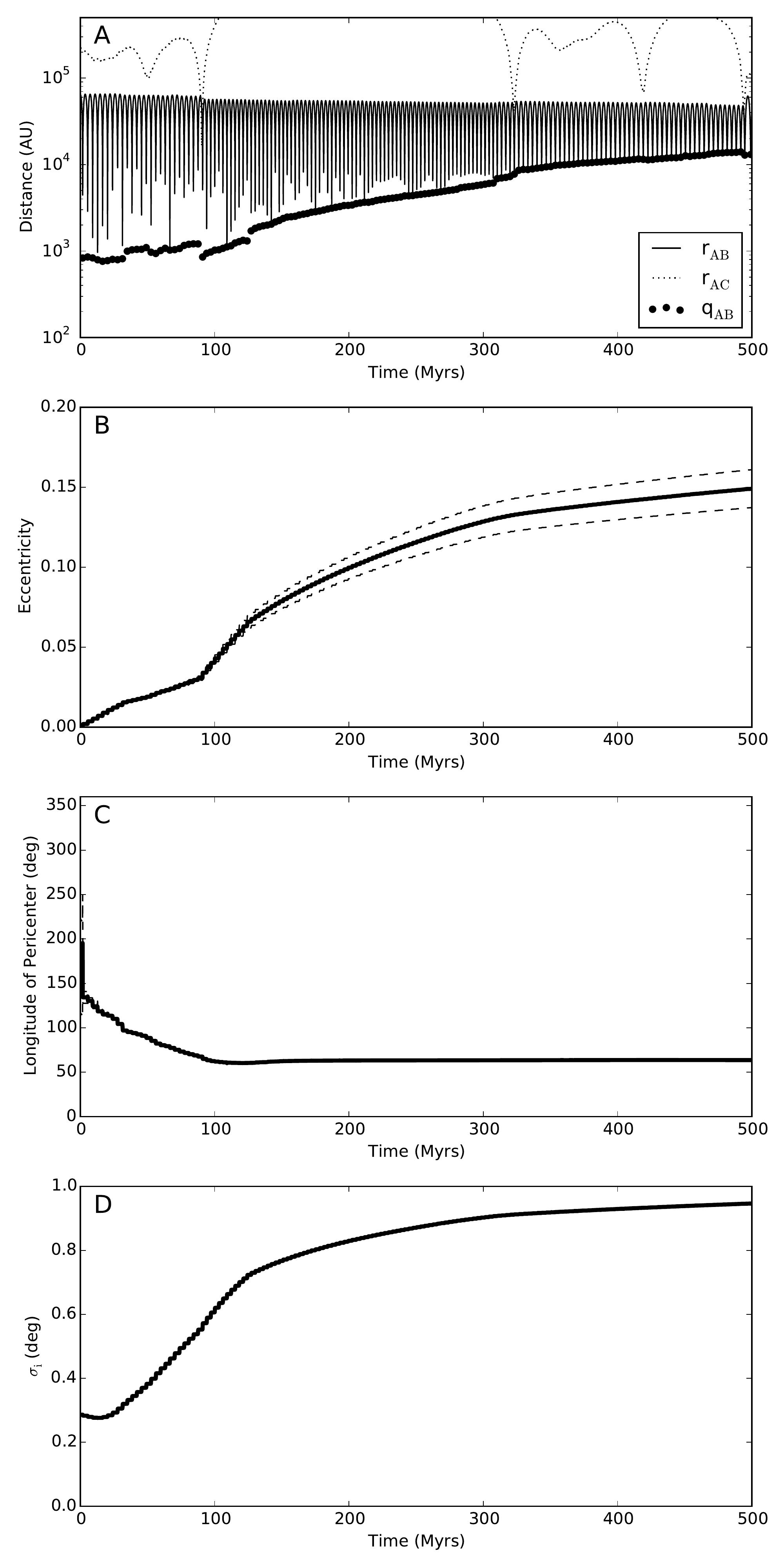}
\caption{{\bf A:} The separation between Fomalhaut A and B ({\it solid line}) as well as between Fomalhaut A and C ({\it dotted line}) is plotted against time for one system. The exact values of each periastron passage inside of 15,000 AU are marked with circular data points. {\bf B:} Median eccentricity of Fomalhaut A's belt as a function of time ({\it solid line}). Dashed lines mark the standard deviation of eccentricity values about the median value. {\bf C:} Median longitude of pericenter of Fomalhaut A's belt as a function of time ({\it solid line}). Dashed lines mark the standard deviation of longitudes about the median longitude. {\bf D:} Standard deviation of the orbital inclinations within Fomalhaut A's belt as a function of time.}
\label{fig:goodexamp}
\end{figure}

On the surface, this result is very encouraging with respect to the hypothesis that Fomalhaut B has driven the eccentric shape of Fomalhaut A's belt. Indeed, when we isolate the 8 systems with median eccentricities between 0.05 and 0.25 (those most comparable to the observed belt's eccentricity of 0.11) we find cases where the observed belt morphology is very closely matched. Perhaps the closest match is shown in Figure \ref{fig:goodexamp}. In this case, Fomalhaut B is already started on a very eccentric orbit, and its initial periastron is 830 AU. For the first 90 Myrs, Fomalhaut B's periastron begins slowly moving outward. This first set of many distant periastron passages drives the median eccentricity of Fomalhaut A's belt to $\sim$0.03. However, at 90 Myrs Fomalhaut B and C undergo a close encounter, which drives B's periastron back in to $\sim$850 AU. In addition, the encounter reorients B's orbital inclination from nearly retrograde to nearly polar relative to the plane of Fomalhaut A's belt. In this new orientation, B's periastron passages have a stronger effect on the belt of A, and the belt's median eccentricity is driven up more rapidly. As the Galactic tide slowly pulls the periastron of B away from A over the last 500 Myrs, the perturbing strength of B's periastron passages slowly diminish. This very extended sequence of moderately close periastron passage yields a final median belt eccentricity of 0.15.

However, replicating the observed features of the Fomalhaut A belt requires more than just generating a median eccentricity that matches the observed belt's eccentricity. We have also seen in the previous section that the best fit to ALMA observations of the belt suggest that the standard deviation of the belt's eccentricity should be below $\sim$43\% of the median eccentricity. In Figure \ref{fig:goodexamp}B we also plot how the standard deviation of the belt evolves with time, and we see it is 0.024 at the end of the simulation, or just 16\% of the median eccentricity of 0.15. 

In addition, the individual orbits of Fomalhaut A's belt must also be apsidally aligned in order for the belt to take a coherent elliptical shape. The best fit of \citet{mac17} suggests that the standard deviation of the longitude of pericenter should be between 6 and 43$^{\circ}$. We test for this requirement in Figure \ref{fig:goodexamp}C where we plot the standard deviation of all the longitudes of pericenter (as well as the median value) as a function of time. (These are calculated relative to the belt's mean orbital plane.) For a uniform distribution of longitudes between 0 and 360$^{\circ}$ we expect a standard deviation of $\sim$104$^{\circ}$, and indeed, our belt initially has a standard deviation of 106$^{\circ}$, consistent with the random distribution of longitudes we assign it initially. However, after just a few periastron passages perturbations from Fomalhaut B have apsidally aligned the belt's orbits, and the belt finishes the simulation with a standard deviation in longitude of pericenter of only 0.16$^{\circ}$. Thus, this system is actually much more apsidally aligned than the real system! It is not clear that this should be considered a failure, for we will see that our simulated belts fail to have the observed degree apsidal alignment more often than not.

Finally, the real belt of Fomalhaut A must have a dynamically cold distribution of orbital inclinations. Observations have constrained the belt objects' mutual inclinations to be $\lesssim$1$^{\circ}$ \citep{bol12}. In Figure \ref{fig:goodexamp}D, we plot how the standard deviation of the belt's inclinations (relative to its mean plane) vary with time. We see that this belt begins with a very small inclination dispersion as specified by our initial conditions. As Fomalhaut B's periastron passages excite the belt's eccentricity, they also excite its inclinations. However, this excitation is modest. At the end of the simulation, the belt's inclinations still have a standard deviation of just below 1$^{\circ}$, which is again consistent with observations. 

\begin{figure} 
\centering
\includegraphics[scale=0.43]{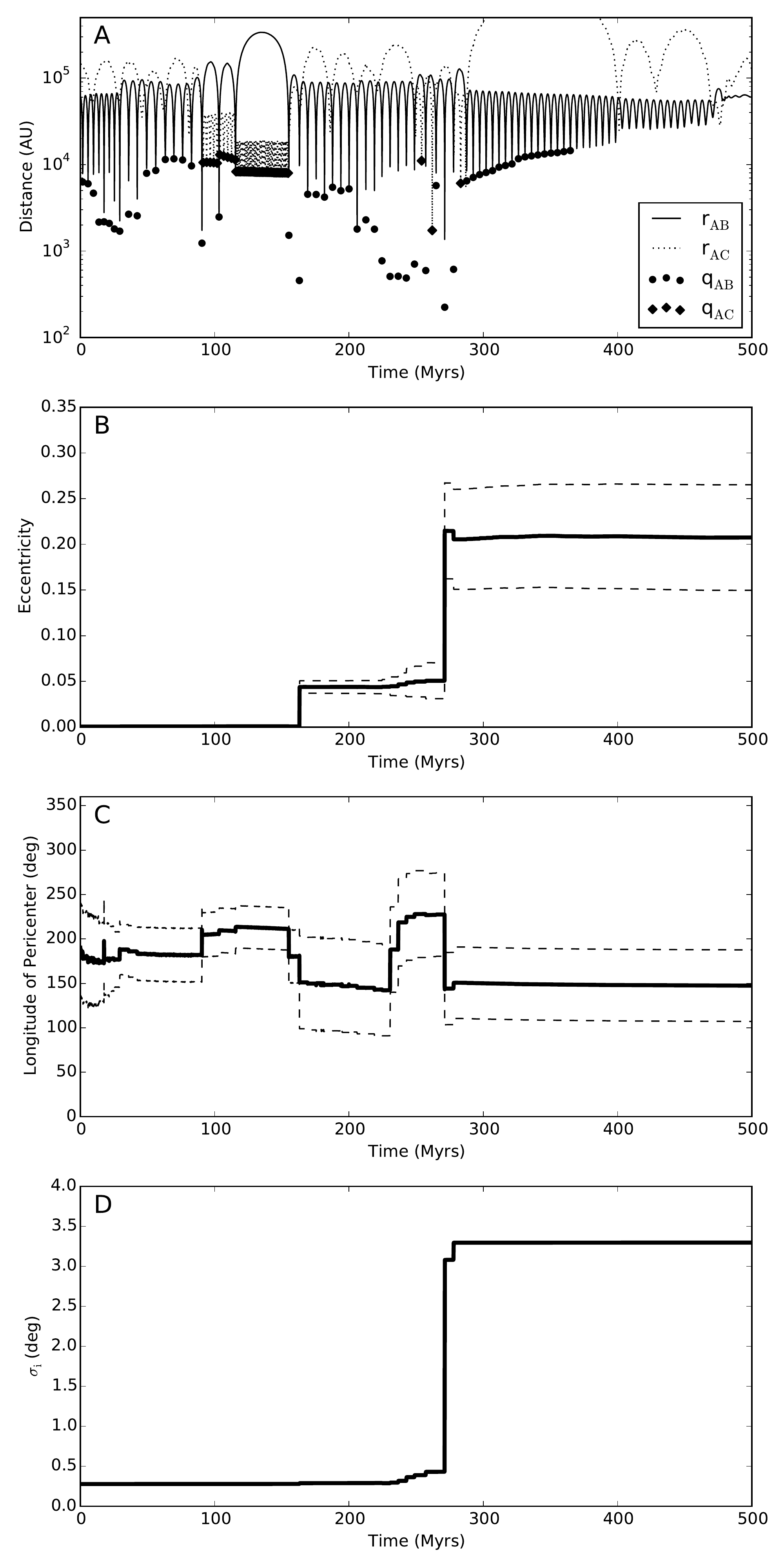}
\caption{Panels, lines and symbols are the same as the previous figure but for a different simulation. In panel {\bf A}, exact periastron passage values of Fomalhaut C that are within 15,000 AU of Fomalhaut A are marked with diamonds.}
\label{fig:badexamp}
\end{figure}

Although the system in Figure \ref{fig:goodexamp} provides an excellent match to the observed Fomalhaut system, the stellar orbital evolution exhibited here is rather uncommon. In this particular simulation, Fomalhaut B begins with a small periastron, and it slowly drifts away from Fomalhaut A over hundreds of Myrs. Moreover, instead of exciting the belt's eccentricity with one or a handful of very close ($\lesssim400$ AU) periastron passages, Fomalhaut B drives up the belt's eccentricity through many moderately close ($\sim$1000 AU) periastron passages. 

It is therefore instructive to study a simulation that excites Fomalhaut A's belt through a smaller number of more powerful periastron passages. One such example is shown in Figure \ref{fig:badexamp}. In this system, Fomalhaut B begins on an orbit with a periastron of 6300 AU. As a result, Fomalhaut A's belt remains nearly circular for the first 100-200 Myrs of the simulation (see Figure \ref{fig:badexamp}B). It is not until interactions between Fomalhaut B and C drive B's periastron down to 460 AU at $t=160$ Myrs that the belt's eccentricity is excited to 0.05. It remains near this value for about 100 Myrs until B's periastron is then driven down to 220 AU. This periastron passage is then powerful enough to drive the belt's eccentricity up to 0.21. The Galactic tide then steadily pulls Fomalhaut B's periastron out to tens of thousands of AU, and the belt's eccentricity is largely unchanged after this. 

One immediately notices in Figure \ref{fig:badexamp}B, that this simulated belt has a much larger eccentricity dispersion than the one shown in Figure \ref{fig:goodexamp}. At $t=500$ Myrs, the eccentricity standard deviation is 0.12, or over half the value of the median eccentricity. Thus, this simulation fails to produce a belt where the eccentricity dispersion is much less than the typical eccentricity. 

It also likely fails with respect to apsidal alignment. In Figure \ref{fig:badexamp}C, we once again plot the standard deviation in the belt's longitudes of pericenter. Although the standard deviation falls by roughly half to $\sim$48$^{\circ}$ over the first 160 Myrs, the first very close periastron passages in the simulation strongly degrade the alignment, and the belt finishes with a standard deviation of 81$^{\circ}$, substantially larger than the range predicted by the observations of \citet{mac17}. 

Finally, while not as striking as the eccentricity dispersion and apsidal misalignment, this belt is also less coplanar than the observed system. Figure \ref{fig:badexamp}D shows that the belt maintains a very coplanar configuration until $t\simeq270$ Myrs. At this point the very close periastron passages of Fomalhaut B increase the spread of inclinations in the belt, so that the standard deviation is above $3^{\circ}$ at the end of the simulation.

\begin{figure} 
\centering
\includegraphics[scale=0.43]{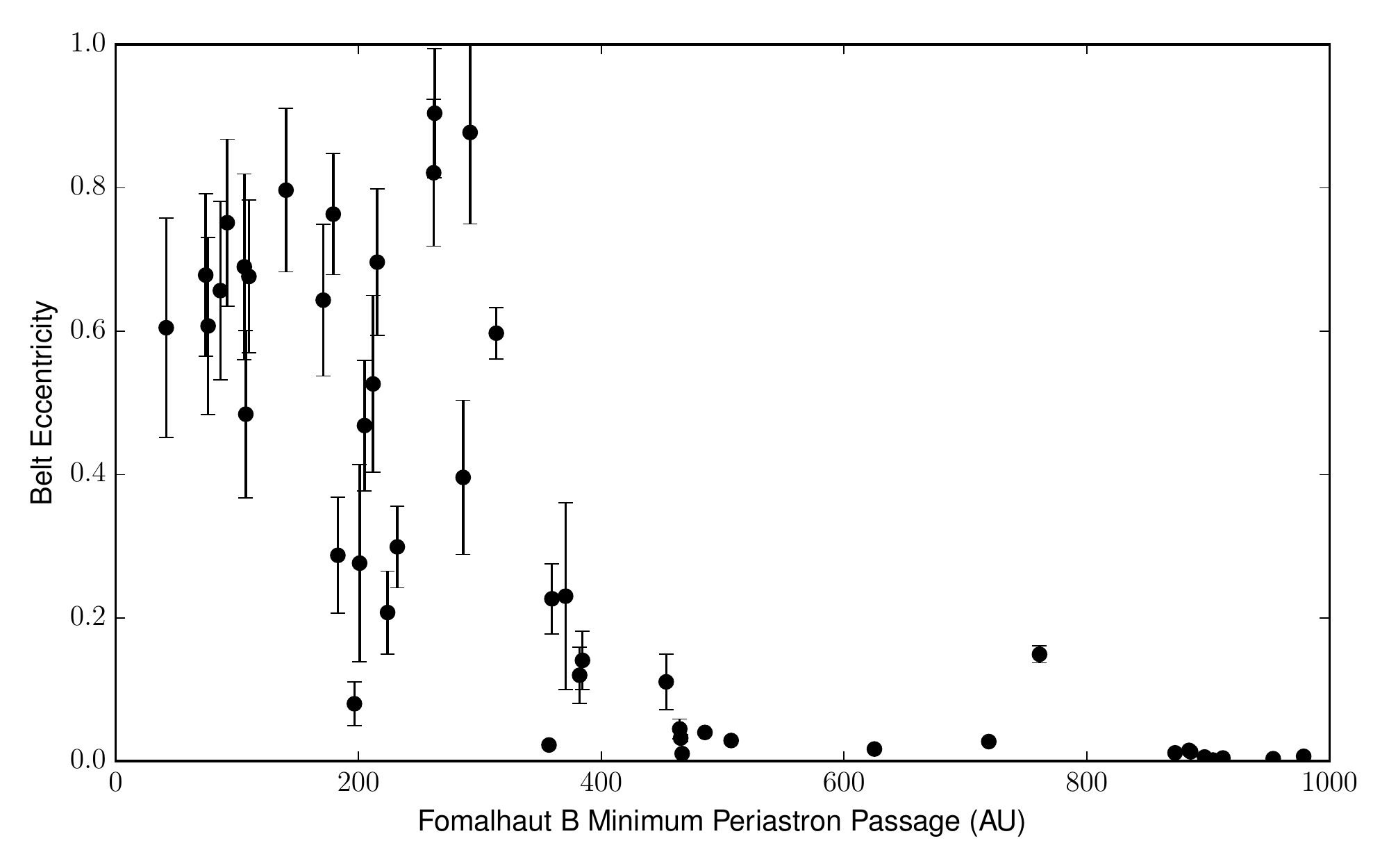}
\caption{The eccentricity of each belt of Fomalhaut A at $t=500$ Myrs is plotted against the minimum periastron passage of Fomalhaut B in each simulation. Data points mark the median eccentricity, while error bars mark each belt's standard deviation.}
\label{fig:ringeccsummary}
\end{figure}

Anecdotally, Figure \ref{fig:badexamp} suggests that belts excited by a handful of very close periastron passages may possess a much higher eccentricity dispersion than that inferred by observations. Figure \ref{fig:ringeccsummary} confirms this. Here we show each belt's eccentricity as a function of the minimum periastron passage distance Fomalhaut B attained in the simulation. (None of the systems where Fomalhaut B's periastron always stayed above $10^3$ AU yielded significantly excited belts.) We see here that our simulations can be split almost perfectly between those where Fomalhaut B passed within 400 AU of Fomalhaut A and those where it did not. In simulations with a periastron passage inside 400 AU, there is only one belt that managed to retain a median eccentricity below 0.05. In contrast, all but 2 belts have median eccentricities below 0.05 if there is no periastron passage within 400 AU (the evolution of one of which is shown in Figure \ref{fig:goodexamp}).

In Figure \ref{fig:ringeccsummary}, we see that a very close periastron passage of Fomalhaut B within 400 AU is by far the most common way to excite the eccentricities within Fomalhaut A's belt. One of the only exceptions to this is the simulation from Figure \ref{fig:goodexamp} that has a minimum periastron passage of $\sim$750 AU. Examining the error bars in Figure \ref{fig:ringeccsummary}, we also see that periastron passages inside 400 AU greatly increase the standard deviation of eccentricities within the belt as well. Although close periastron passages of Fomalhaut B are not unusual and they can significantly excite eccentricities within Fomalhaut A's belt, they nearly always leave the belt with a large spread of eccentricities.

\begin{figure} 
\centering
\includegraphics[scale=0.43]{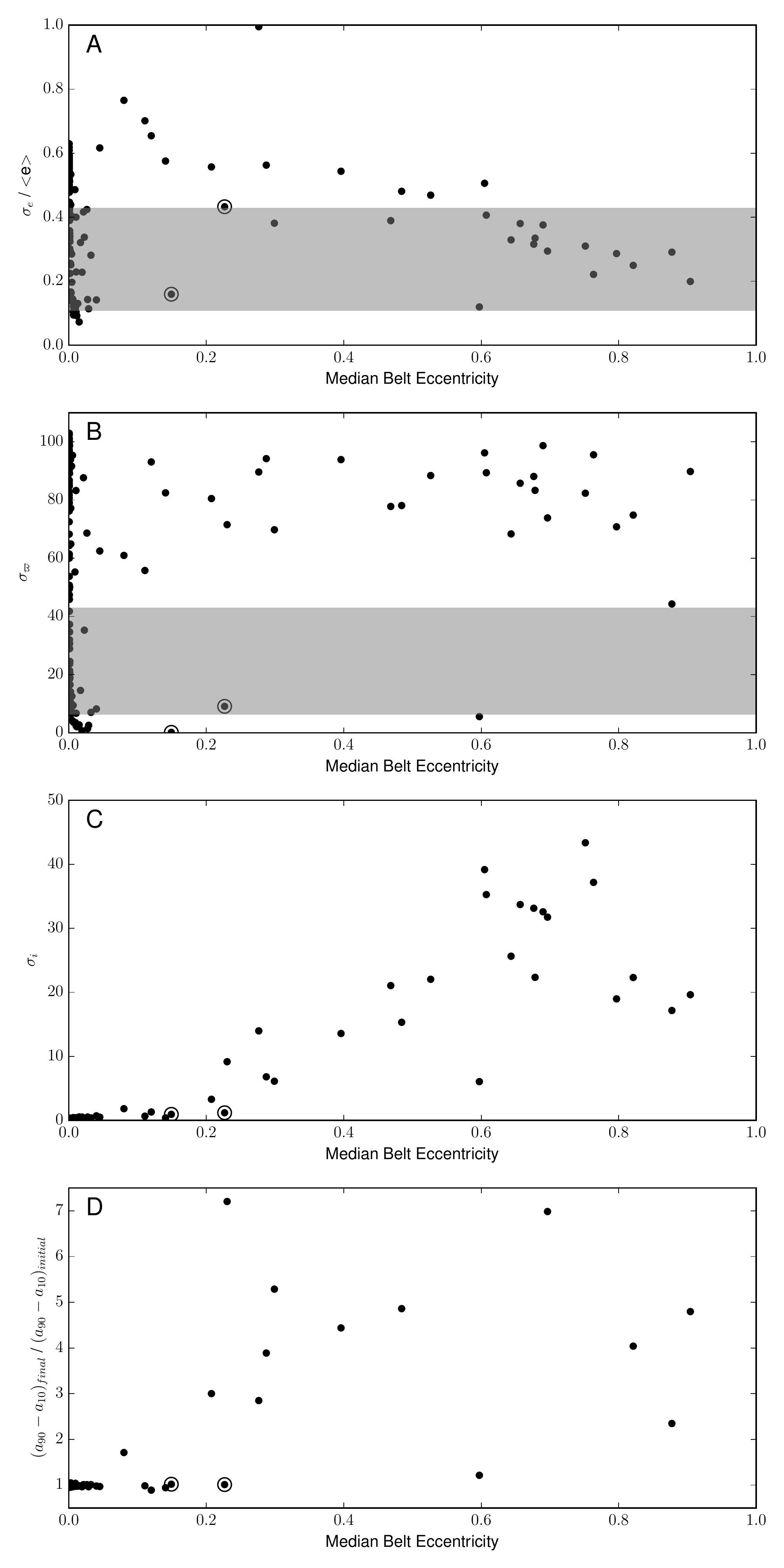}
\caption{{\bf A:} Ratio of the eccentricity standard deviation to the median eccentricity vs the median eccentricity of each of our belts around Fomalhaut A at $t=500$ Myrs. Shaded area marks the range of ratio values corresponding to the best fit to ALMA observations of Fomalhaut's belt \citep{mac17}. {\bf B:} Standard deviation of the longitudes of pericenter for each Fomalhaut A belt plotted against its median eccentricity. Shaded area marks the range of standard deviation values corresponding to the best fit to ALMA observations of Fomalhaut's belt \citep{mac17}. {\bf C:} Standard deviation of the orbital inclinations for each Fomalhaut A belt plotted against its median eccentricity. {\bf D:} Ratio of the semimajor axis range encompassing the middle 80\% of belt particles at the end of each simulation to the beginning of each simulation. }
\label{fig:ringallsummary}
\end{figure}

This is further borne out in Figure \ref{fig:ringallsummary}A. Here we see that belts with median eccentricities near the observed value usually have a standard deviation within a factor of $\sim$2 of the median. Smaller ratios of standard deviation to the median value do exist, but they typically involve belts that are nearly circular ($e<.05$) or belts that are extremely eccentric ($e \gtrsim0.6$). 

However, even though close periastron passages of Fomalhaut B significantly excite the spread of eccentricities, matches to the observed belt may still be attained. Out of our 8 simulated belts with the most promising median eccentricities ($0.05<e<0.25$), we find 2 belts with eccentricity spreads comparable to the observed belt. One is the simulation detailed in Figure \ref{fig:goodexamp}, and the other is a belt whose ratio of eccentricity standard deviation to median eccentricity is 0.43, right at the upper limit predicted in the best-fit model of \citet{mac17}. Unlike our simulation from Figure \ref{fig:goodexamp}, this system does experience very close periastron passages of Fomalhaut B. Locating this belt in Figure \ref{fig:ringeccsummary} (median eccentricity of 0.23), we see that Fomalhaut B's minimum periastron passage is 360 AU. Thus, it is possible to produce a belt via strong perturbations from Fomalhaut B whose eccentricity spread is similar to the observed belt.

In Figure \ref{fig:ringallsummary}B we see that, more often than not, very close periastron passages of Fomalhaut B yield a belt that is less apsidally aligned than then observed belt. Nevertheless, if we again only study the 8 belts with median eccentricities between 0.05 and 0.25, we find that 2 of them are at least as apsidally aligned as the observed belt. Moreover, these are the same 2 belts that provided the best matches to the observed belt's eccentricity dispersion. 

In Figure \ref{fig:ringallsummary}C we look at the degree of coplanarity in all of our simulated belts. Unlike apsidal alignment, our simulated belts are largely coplanar for median eccentricities near the observed eccentricity. Although there are belts with standard deviations in inclination of 10--50$^{\circ}$ these almost all occur when the median eccentricity is driven above 0.25. For the 8 cases most similar to the observed eccentricity ($0.05<e<0.25$), we only have 2 simulated belts with an inclination standard deviation above 2$^{\circ}$. Thus, the observed belt's degree of coplanarity seems to be a relatively easy feature to replicate in our simulations. 

Finally, while it is not obvious how perturbations from Fomalhaut B could generate the observed belts sharp edges, these sharp edges must be able to survive close periastron passages of Fomalhaut B for our mechanism to be viable. We study changes to the belt's edges in Figure \ref{fig:ringallsummary}D. Here we compare the range of semimajor axes that enclose the middle 80\% of all ring particles at the beginning of each simulation with this range of semimajor axes at the end of each simulation. We see that for belts that have only been moderately excited to median eccentricities below $\sim$0.2, this range of semimajor axes typically does not change much throughout a simulation. However, perturbations driving the median eccentricity above $\sim$0.2 typically also cause a dramatic diffusion in the range of semimajor axes confining the belt. Typically, this range increases by at least a couple hundred percent. It should be noted that our two best matches to the observed belt in terms of eccentricity dispersion and apsidal alignment are also consistent with the narrow semimajor axis confinement of Fomalhaut A's belt. Over the course of 500 Myrs of evolution the semimajor axes confining the middle 80\% of these rings orbits grows by just 1--2\% in each simulation. Thus, these two simulations satisfy every major orbital constraint given by observations of the real belt.

\subsection{Distinguishing Scenarios}

Because our work shows it is possible for a low periastron passage of Fomalhaut B to generate the morphology of Fomalhaut A's belt, we now turn our attention toward the differences between the various scenarios invoked in the literature to explain the belt's morphology. Although perturbations from nearby planets can explain the morphology, a definitive planetary object has yet to be detected around Fomalhaut A. Clearly, if a planet or planets are detected on orbits consistent with the driving of an eccentric belt, this explanation for the belt morphology becomes the favored one.

To date, only our work here and that of \citet{shan14} explore the possibility that the belt morphology is a consequence of perturbations from Fomalhaut A's stellar companions. The main difference between these two mechanisms is that \citet{shan14} predict that the belt's morphology is generated during a reshuffling of the stellar companion hierarchy that usually results in Fomalhaut C's ejection from the system. Meanwhile, in our work the belt's eccentricity is nearly always attained when galactic perturbations drive the periastron of Fomalhaut B to very low values. Although our stellar systems can also go unstable, this is not a requirement by any means. This can be seen when we look at the recent dynamical histories of our systems whose final stellar separations are within $\pm$50\% of the observed separations. In $\sim$80\% of these systems, the last apastron passage of Fomalhaut B with respect to Fomalhaut A is smaller than the last periastron passage of Fomalhaut C with respect to Fomalhaut A. This indicates that most of our systems that match the observed stellar separations have non-crossing stellar orbits, which we take as a proxy for systems where stars B and C do not strongly interact. On the other hand, \citet{shan14} report that less than 20\% of their matching systems finish with Fomalhaut B and C on non-interacting orbits. Thus, a very precise measurement of the stellar companions' velocities may be one way to help distinguish our scenario from that of \citet{shan14}. 

In addition, the degree of apsidal alignment may be another key to distinguishing between a disk morphology generated by close periastron passages of Fomalhaut B (the work presented here) and the stellar instability process described in \citet{shan14}. In our work, we find that nearly all of our eccentric belts have a noisy apsidal alignment. Out of all of our simulated belts with median eccentricities above 0.05, we only find one case whose standard deviation in longitude of pericenter is below 1$^{\circ}$. Meanwhile, an examination of the \citet{shan14} results finds that they predict standard deviations near 0 unless the median belt eccentricity exceeds $\sim$0.3. If future observations rule out a standard deviation in longitude of pericenter of 0, this would favor our scenario. On the other hand, an extremely aligned belt is easier to produce in the \citet{shan14} mechanism. 

One final potential way to differentiate between our mechanism and that of \citet{shan14} involves the disk of Fomalhaut C. Because star C is perturbing the Fomalhaut A belt in the \citet{shan14} case, star C must pass near star A. Such close passages are very likely to excite the eccentricities of bodies in the Fomalhaut C disk. Meanwhile, in our mechanism there is no requirement for star C to play a substantial dynamical role, and less than 20\% of our system histories have star C pass within 2000 AU of star A. If future observations reveal an eccentric disk about Fomalhaut C, this would favor the \citet{shan14} mechanism over ours.

All of the distinctions between our mechanism and that of \citet{shan14} that are discussed above rely on differing general tendencies of the two mechanisms rather than phenomena that are prohibited in one and constantly observed in the other. Thus, to confidently differentiate between these two mechanisms likely requires acquiring a combination of new observational constraints on the Fomalhaut system.

\section{Conclusions}

In this work, we construct a new symplectic integration algorithm that can accurately and efficiently model the orbital dynamics of a planetary system as it evolves within a stellar system or 3 or more stars. Using this code, we perform 2000 simulations of the stellar dynamical evolution of the Fomalhaut triple star system as it has been perturbed by its local galactic environment over the past 500 Myrs. For the 135 dynamical histories that yield stellar configurations similar to the system's observed state, we then simulate the evolution of a belt of small planetesimals about Fomalhaut A as the stellar system progresses through its dynamical evolution. 

From the results of these simulations we can make a number of conclusions. First, present-day separations of Fomalhaut A, B, and C could have been maintained over the system's $\sim$500-Myr history. \citet{shan14} suggest that the currently observed state of Fomalhaut could be a transition phase as Fomalhaut C is being ejected from the stellar system. However, when we begin our simulations with stellar orbits that are consistent with the modern separations, we find that after 500 Myrs of evolution, $\sim$7\% of our systems still have separations comparable to the observed values. This fraction of systems is substantially higher than that of \citet{shan14}. Moreover, over half of our systems have possessed stellar separations consistent with the observed values at some point during their final 100 Myrs of integration. Thus, we find it very plausible that the typical stellar separations and hierarchy of Fomalhaut have not changed markedly over the system's lifetime.

Our finding that many initial states can remain stable for the age of Fomalhaut is not obvious. Because the stellar separations of Fomalhaut are $\sim$10$^{4-5}$ AU, the stellar orbits are strongly affected by the Galactic tide as well as other passing field stars. As a result, the stellar eccentricities and inclinations are continually changing. This often can lead to close encounters between stellar companions as they pass through phases of very low periastron. It is nearly inevitable that Fomalhaut B and C make at least one close periastron passage of Fomalhaut A during the course of 500 Myrs of evolution. The median values for the minimum periastron passages of B and C with respect to Fomalhaut A are 2500 and 9800 AU, respectively.

Such periastron passages can significantly alter the morphology of a belt of planetesimals in orbit around Fomalhaut A. If we begin with a nearly circular and coplanar belt positioned at the distance of Fomalhaut A's observed belt, we find that periastron passages of Fomalhaut B commonly excite the planetesimals' orbital eccentricities. The minimum periastron passage typically necessary for significant excitation is $\sim$400 AU. If this is attained, eccentricities comparable to or well in excess of the observed belt's eccentricity are nearly always attained. Such close periastron passages occur in about 1/4 of our simulated systems.

Often times, the eccentric belts generated in our simulations have a spread of orbital eccentricities larger than that inferred by observations of Fomalhaut A's real belt \citep{mac17}. Similarly, our simulated belts are typically less apsidally aligned than the observed belt. Nevertheless, if we restrict our analysis to the 8 simulated belts with median eccentricities most comparable to the observed belt eccentricity, we find that 25\% of them have eccentricity dispersions consistent with the observed belts and that these systems are at least as apsidally aligned as the observed belt. Moreover, these simulated belts have the same degree of coplanarity and semimajor axis confinement of the observed belt, constraints that we found are more regularly replicated.

Out of our 135 systems that finish with the observed stellar separations, only 2 also fully replicate all the observed features of Fomalhaut A's belt. However, this result should not be used to dismiss stellar perturbations as the cause of the observed belt's morphology. In our simulated systems, there is a $\sim$25\% chance that Fomalhaut B's periastron passages will transform Fomalhaut A's belt into an eccentric state. If this occurs, the diversity of outcomes is very large, and the observed properties of Fomalhaut A's belt fall well within the spectrum of belt morphologies that we predict.

On the other hand, one also cannot confidently state that perturbations from Fomalhaut B are in fact responsible for the observed morphology of Fomalhaut A's belt. It is well-known that such an eccentric belt can be produced from the secular forcing of planets in orbit about Fomalhaut A \citep[e.g., ][]{wyatt99}. Nevertheless, it is not known whether such planets reside in this system. If future observations of Fomalhaut A rule out the existence of planets capable of generating its belt's morphology, the most plausible explanation for Fomalhaut A's eccentric belt is close periastron passages of Fomalhaut B, which has been definitively confirmed as a member of the system \citep{mam12}. 

More broadly speaking, features or asymmetries of planetary belts or disks are often employed as a way to infer the presence of one or more planets within a system \citep{wyatt99}. This approach has already predicted the existence of planets before their direct detection \citep[e.g.][]{lag10}. Nevertheless, our work here shows that very distant stellar companions can, counterintuitively, often deliver strong perturbations to these structures that can mimic the signatures of planets. 

\section{Acknowledgements}

We thank John Chambers and Sean Raymond for useful discussions as well as the anonymous referee whose comments and suggestions greatly improved the quality of our submission. This work was supported by HST Theory grant HST-AR-13898.001-A as well as the University of Oklahoma Physics REU program (NSF Award \#1359417). A.I thanks financial support from FAPESP via proc. 16/19556-7 and 16/12686-2. The computing for this project was performed at the OU Supercomputing Center for Education \& Research (OSCER) at the University of Oklahoma (OU).

\appendix
\section*{Appendix A: Conjugate Momenta}
\renewcommand{\theequation}{A\arabic{equation}}

In this appendix, we use a generating function to derive the canonical conjugate momenta of the coordinates we employ in our new algorithm. The generating function used in \citet{cham02} to relate their integrator coordinates and conjugate momenta is 

\begin{equation} \label{eq:chamgen}
\begin{aligned}
F_g = &-\bm{p}_A \cdot \left(\bm{X}_{A} - \frac{m_B}{m_A + m_B + \sum_j m_j}\bm{X}_B - \frac{\sum_j m_j \bm{X}_j}{m_A + \sum_j m_j}\right) \\\\
&- \sum\limits_i \bm{p}_i \cdot \left( \bm{X}_i + \bm{X}_A  + \frac{m_B}{m_A + m_B + \sum_j m_j} \bm{X}_B - \frac{\sum_j m_j \bm{X}_j}{m_A + \sum_j m_j}\right)\\\\
&- \bm{p}_B \cdot \left( \bm{X}_A + \frac{m_A + \sum_j m_j}{m_A + m_B + \sum_j m_j}\bm{X}_B\right)
\end{aligned}
\end{equation}
To find our new algorithm's coordinate conjugate momentum pairs, we make a simple modification to the \citet{cham02} function as follows:

\begin{equation} \label{eq:mygen}
\begin{aligned}
F_g = &-\bm{p}_A \cdot \left(\bm{X}_{A} - \frac{m_B}{m_A + m_B + \sum\limits_{j=1}^{N_P} m_j}\bm{X}_B - \frac{\sum\limits_{j=1}^{N_P} m_j \bm{X}_j}{m_A + \sum\limits_{j=1}^{N_P} m_j}\right) \\
&- \sum\limits_{i=1}^{N_P} \bm{p}_i \cdot \left( \bm{X}_i + \bm{X}_A  + \frac{m_B}{m_A + m_B + \sum\limits_{j=1}^{N_P} m_j} \bm{X}_B - \frac{\sum\limits_{j=1}^{N_P} m_j \bm{X}_j}{m_A + \sum\limits_{j+1}^{N_P} m_j}\right)\\
&- \bm{p}_B \cdot \left( \bm{X}_A + \frac{m_A + \sum\limits_{j=1}^{N_P} m_j}{m_A + m_B + \sum\limits_{j=1}^{N_P} m_j}\bm{X}_B\right)\\
& - \sum\limits_{i=N_P+1}^{N_P+N_S} \bm{p}_i\cdot \bm{X}_i
\end{aligned}
\end{equation}

To demonstrate that this generating function yields the coordinates and momenta given in Equations \ref{eq:mycoords} and \ref{eq:mymom}, we first invert Equation \ref{eq:mycoords} to express the inertial coordinates in terms of our new algorithm's coordinates:

\begin{equation} \label{eq:invertedcoords}
\begin{aligned}
\bm{x}_A &= \bm{X}_A-\frac{\sum\limits_{i=1}^{N_P}m_i\bm{X}_i}{m_A + \sum\limits_{i=1}^{N_P}m_i} - \frac{m_B \bm{X}_B}{m_A + m_B + \sum\limits_{i=1}^{N_P}m_i},\\\\
\bm{x}_i &= \bm{X}_i + \bm{X}_A-\frac{\sum\limits_{i=1}^{N_P}m_i\bm{X}_i}{m_A + \sum\limits_{i=1}^{N_P}m_i} - \frac{m_B \bm{X}_B}{m_A + m_B + \sum\limits_{i=1}^{N_P}m_i}  \quad\quad\quad {\rm for}~1\leq i\leq N_P,\\\\
\bm{x}_B &=  \bm{X}_A + \frac{m_A + \sum\limits_{i=1}^{N_P}m_i}{m_A + m_B + \sum\limits_{i=1}^{N_P}m_i}\bm{X}_B,\\\\
\bm{x}_i &= \bm{X}_i ~\quad\quad\quad\quad\quad\quad\quad\quad\quad\quad\quad\quad\quad\quad\quad\quad\quad\quad\quad\quad\quad~ {\rm for}~N_P< i\leq N_P+N_S
\end{aligned}
\end{equation}

Using the generating function in Equation \ref{eq:mygen}, we see that our original, inertial coordinates are recovered through

\begin{equation} \label{eq:recoverinert}
x_i = -\frac{\partial F_g}{\partial p_{x,i}}
\end{equation}
Similarly, the conjugate momenta to the coordinates employed in our new algorithm are given by

\begin{equation} \label{eq:getnewmom}
P_{x,i} = -\frac{\partial F_g}{\partial X_{i}}
\end{equation}
which are detailed in Equation \ref{eq:mymom} of the main paper. 

\section*{Appendix B: Testing of Algorithm}
\renewcommand{\thefigure}{B\arabic{figure}}
\setcounter{figure}{0} 

It is important to understand the accuracy limits of our new algorithm. Like other symplectic codes, we expect our simulations to have a small, bounded numerical error under most circumstances. However, we anticipate the quality of our numerical integrations will be degraded during certain instances. Some of these instances are well-studied and result from the inability of democratic heliocentric coordinates to accurately model close pericenter passages between planetary objects and the primary star \citep{levdun00}. In these cases, $H_{\rm jump}$ becomes large relative to $H_{\rm Kep}$ and can no longer be treated as a perturbation on a near Keplerian trajectory. 

However, other cases of numerical degradation are unique to our new algorithm. In particular, our code has no means of handling close encounters between the secondary star and additional stars. When these two classes of bodies approach one another the second-to-last set of terms in $H_{\rm int}$ becomes large and degrades the integration of the binary and the other star involved in the encounter. Another case of concern involves close encounters between planetary bodies and either the secondary star or the additional stars. During these situations, the third and fourth sets of terms in $H_{\rm int}$ can become large and degrade the integration of the bodies involved in the close encounter {\it as well as} other bodies' integrations due to the presence of indirect accelerations in Equations \ref{eq:accelpl} and \ref{eq:accelbin}. Finally, close encounters between passing stars and the primary star can cause the last term in $H_{\rm int}$ to greatly increase as well and degrade integration quality.

Thus, it is important to understand how closely objects of various classes can pass within one another before the quality of our simulations is significantly compromised. Adequately testing our code is challenging because it is designed to model complex heirarchical systems whose dynamics have no analytical solution to serve as a reference. However, we have settled upon simulating various manifestations of the circular restricted three-body problem and monitoring the conservation of the Jacobi Constant. The problem is heirarchical in nature, just as our system of interest, Fomalhaut, is. 

With the Fomalhaut system in mind, we integrate 50 simulations of a 3-body system consisting of a 1.92 M$_{\sun}$ primary (star A), a 0.73 M$_{\sun}$ secondary (star B), and a massless test body (star C, which can be thought of as a Fomalhaut C pseudo-analog, as the restricted 3-body problem requires one nearly massless body). In our initial conditions, the secondary is placed in a circular orbit about the primary with a semimajor axis of 500 AU, while the test body has an initial semimajor axis of 1000 AU, an eccentricity of 0.3, and a random inclination relative to the secondary's orbital plane. The test body's initial orbit is chosen so that it will quickly undergo strong interactions with the secondary. The initial separations of our bodies are much smaller than the separations in the Fomalhaut system. Again, this is done to ensure that close encounters between the bodies occur regularly. Our 50 systems are integrated until the test body is ejected ($r>10^5$ AU and positive energy with respect to the center of mass). Since we do not know the true stellar orbits of the Fomalhaut system, and it may be in the process of going unstable, it is important for us to accurately integrate such a system completely through an instability. In this set of simulations the secondary star is integrated using the wide binary star coordinate scheme of \citet{cham02}, and the test body is integrated via $T+V$ in the inertial coordinate extension that we have designed. 

\begin{figure}
\centering
\includegraphics[scale=0.5]{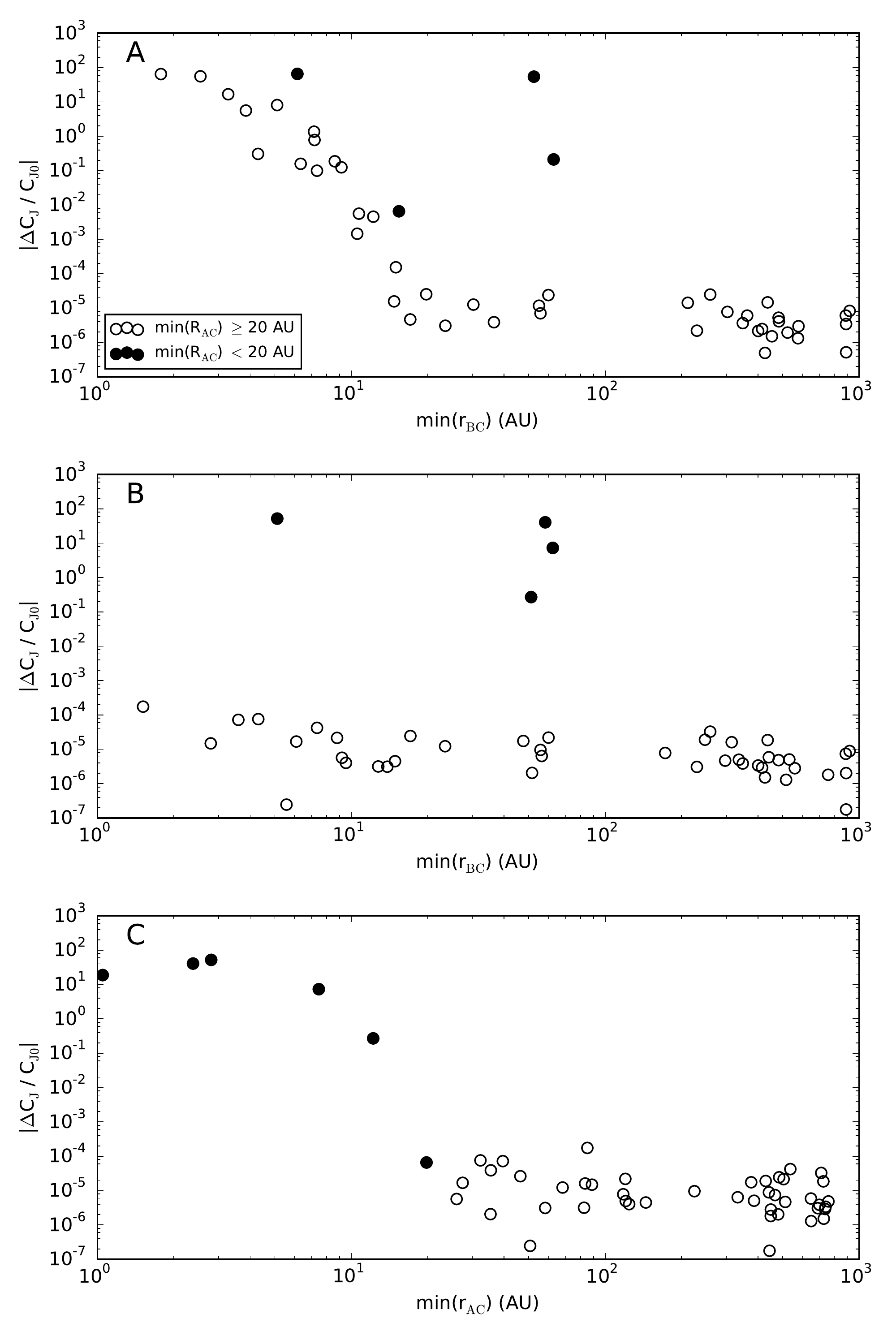}
\caption{{\bf A}: Fractional change in Jacobi Constant between the beginning and end of 50 simulations of a 1.92 M$_{\sun}$ star (body A), a 0.78 M$_{\sun}$ star (body B), and a massless test body (body C). This is plotted against the minimum approach distance of bodies B and C. The 0.78 M$_{\sun}$ star is integrated with the binary coordinate scheme of \citet{cham02}, while the test body is integrated in inertial coordinates. {\bf B}: Fractional change in Jacobi Constant vs the minimum approach distance of bodies B and C. This is for the same 50 simulations from panel A, except they are rerun with both bodies B and C integrated in intertial coordinates. {\bf C}: Fractional change in Jacobi Constant vs the minimum approach distance of bodies A and C for the same simulations from panel B. }
\label{fig:secondaryerr}
\end{figure}

In Figure \ref{fig:secondaryerr}A the numerical error of our systems' Jacobi constant is plotted against the minimum approach distance attained in each system between the secondary star and the massless body. Here we see that there is a clear dependence of the degree to which the Jacobi Constant is conserved on the closest encounter that the secondary has with the test body (tertiary). The Jacobi Constant error begins to significantly increase for encounter distances below $\sim$20 AU, and unfortunately, this represents 40\% of our simulated systems. In addition, we see that there are other instances where the secondary and tertiary remain far from one another and the Jacobi Constant conservation is still poor. In these cases the source of the error is a close encounter between the primary and the tertiary. Again, when the primary and tertiary pass within $\sim$20 AU of one another we find significant increases in numerical error. 

Figure \ref{fig:secondaryerr}A raises a conundrum. Although our code seems accurate as long as stellar companions remain far from one another, encounters between the secondary and tertiary are common, and we have no special routine built into our code to handle such encounters. The reason for this is that the Hamiltonian partition we have chosen assumes that the secondary is on a near-Keplerian orbit about the primary and its planets, while this assumption is dropped for the tertiary's trajectory. This approach enables us to accurately integrate the secondary's orbit to lower pericenter than other stars, but as we see in Figure \ref{fig:secondaryerr}A close encounters between the primary and its stellar companions are less common than encounters between the two orbiting stars. Thus, our integrations may be improved if we abandon the assumption of near-Keplerian motion for the secondary and treat it as another additional star. For this setup, the Hamiltonian described in Equation \ref{eq:Hpart} would have $m_B=0$ and $N_S=2$. 

With this in mind we reintegrate our 50 systems through dissolution with the secondary also handled with a simple $T+V$ integration. Again, we plot the error in Jacobi Constant as a function of the minimum approach distance between the stars B and C in Figure \ref{fig:secondaryerr}B. This time we see that numerical error has no dependence on this approach distance, since our code accurately handles close encounters between these stars with a Bulirsch-Stoer routine. The remaining 4 simulations that have poor Jacobi Constant conservation all have encounters between A and C inside 20 AU. When we replot numerical error as a function of the closest approach distance between stars A and C in Figure \ref{fig:secondaryerr}C, we see a clear dependence on this parameter. This particular source of error is less concerning to us because of the nature of our particular study. We are interested in whether Fomalhaut A's stellar companion has altered its debris belt. This belt, which sits over 100 AU from Fomalhaut A, would likely be obliterated if a stellar companion passed within 20 AU of A. Although the accuracy of the integration is compromised after this periastron passage, we will not be particularly interested in the evolution of the system after this point. Thus, treating the binary as just another additional star gives us an integration scheme that models Fomalhaut's stellar dynamics well enough to study their effects on Fomalhaut A's belt.

\begin{figure}
\centering
\includegraphics[scale=0.5]{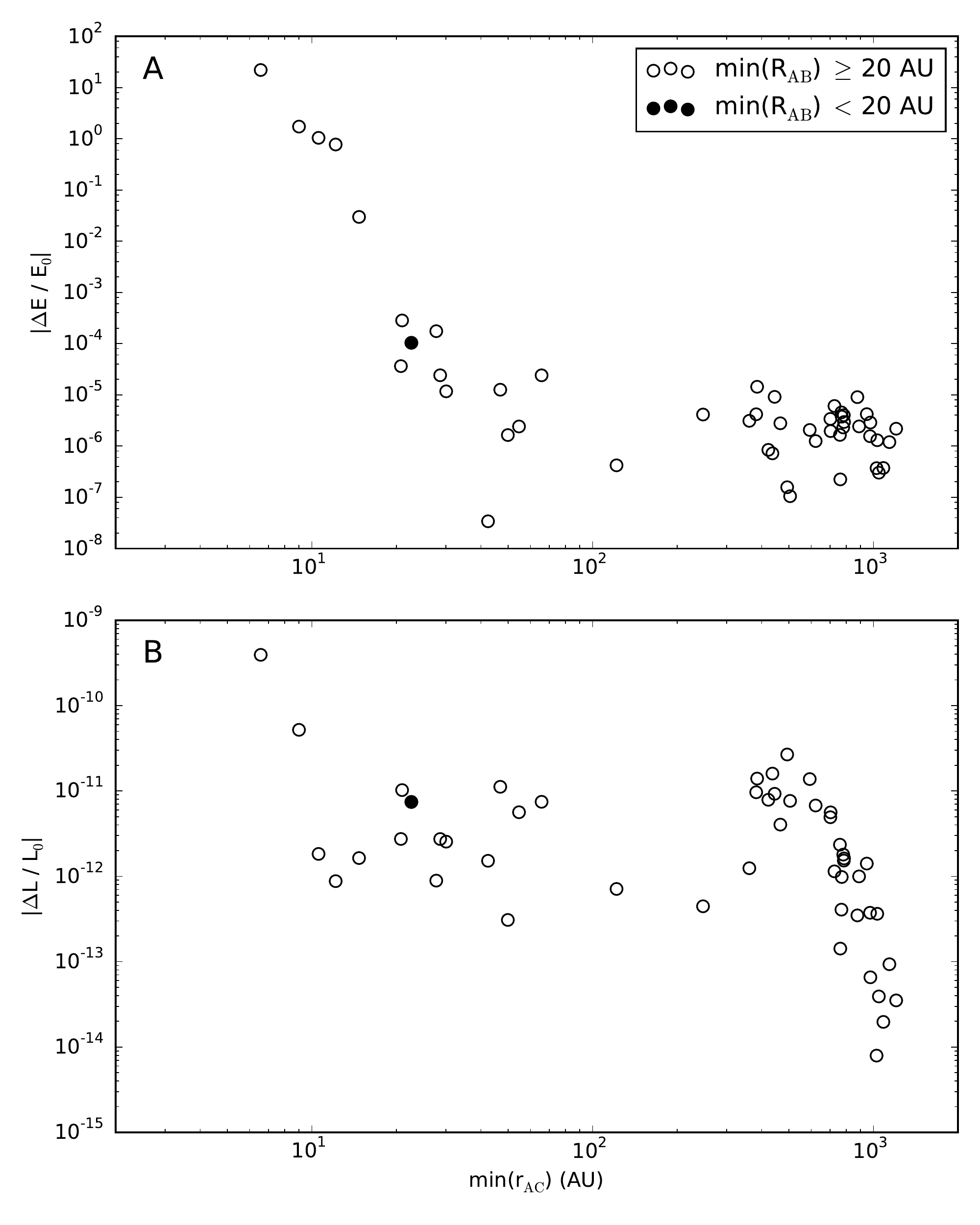}
\caption{{\bf A}: Fractional change in energy between the beginning and end of 50 simulations of a 1.92 M$_{\sun}$ star (body A), a 0.73 M$_{\sun}$ star (body B), and a 0.18 M$_{\sun}$ star (body C). This is plotted against the minimum approach distance of bodies A and C. Data points from simulations in which bodies B and C have an approach within 20 AU are filled. {\bf B}: Fractional change in angular momentum for the same set of simulations.  Data points from simulations in which bodies B and C have an approach within 20 AU are again filled. }
\label{fig:allinerterr}
\end{figure}

Up to now, we have only measured the error in the conservation of the Jacobi Constant. We can also characterize our code's conservation of energy and angular momentum. To do this, we repeat our integrations one more time, except with star C having the actual mass of Fomalhaut C of 0.18 M$_{\sun}$. As with the previous set of integrations, both stars are integrated with a $T+V$ scheme in inertial coordinates. The numerical energy error and angular momentum error for these integrations are shown as a function of the closest approach distance between A and C in Figure \ref{fig:allinerterr}. As with Jacobi Constant, we see that approaches between star A and C that are less than $\sim$20 AU significantly increase the energy error in our simulations. Although angular momentum error is degraded for the closest approaches between A and C, it is still conserved to better than 1 part in 10$^9$. We also expect that close approaches between stars A and B will increase numerical error, but we only have one integration where A and B come within 20 AU, and this system still has a fractional energy error of just 10$^{-4}$. 

While we have demonstrated that the stellar interactions of our new simulation code are accurate enough for our purposes, we have yet to study the error associated with close approaches between stars and planetary bodies orbiting the primary. As with our stellar encounters study, we again set up a version of the circular restricted 3-body problem to characterize this error. In this setup, we place a 0.73 M$_{\sun}$ Star B in a circular orbit with a semimajor axis of 150 AU about Star A whose mass is set to 1.92 M$_{\sun}$. Now we place one more test body on a circular orbit at 120 AU (near the position of Fomalhaut A's ring), which we will still refer to as body C here. Unlike our prior simulations, body C will now be integrated within democratic heliocentric coordinates using a mixed variable symplectic routine, while star B will be integrated with a $T+V$ approach within inertial coordinates. Because of its proximity to Star B, the orbit of C is inherently unstable, and during the instability close approaches between B and C (as well as A and C) are likely to ensue, allowing us to study the numerical error associated with such events. 

\begin{figure}
\centering
\includegraphics[scale=0.5]{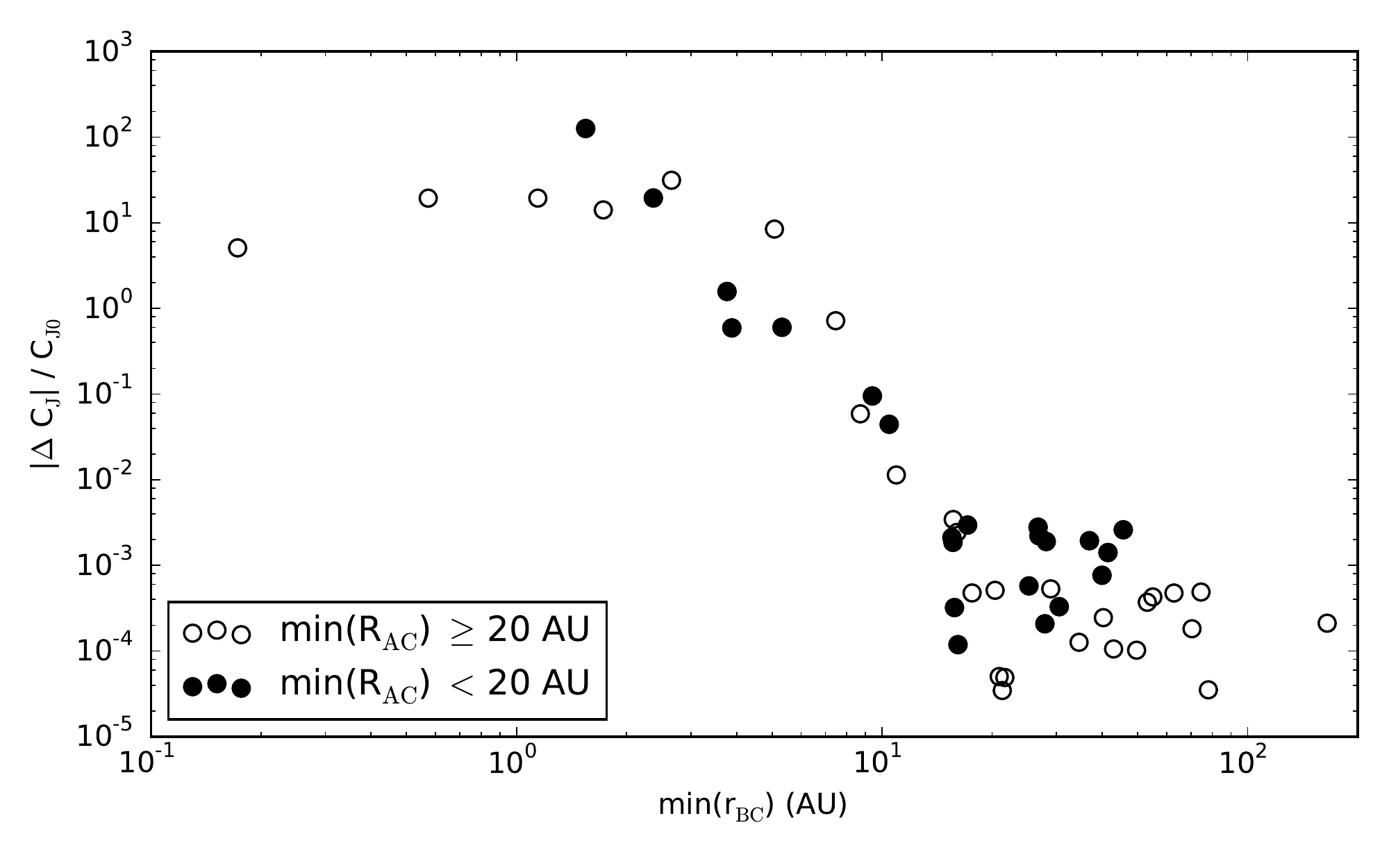}
\caption{Fractional change in Jacobi Constant between the beginning and end of 50 simulations of a 1.92 M$_{\sun}$ star (body A), a 0.78 M$_{\sun}$ star (body B), and a massless test body (body C). This is plotted against the minimum approach distance of bodies B and C. The 0.78 M$_{\sun}$ star is integrated in inertial coordinates, while the test body is integrated in democratic heliocentric coordinates. Simulations in which bodies A and C approach each other within 20 AU are marked with filled symbols.}
\label{fig:starplerr}
\end{figure}

We integrate 50 such systems until body C is ejected from each system. The error in Jacobi Constant of these systems is plotted against the minimum BC approach distance in Figure \ref{fig:starplerr}. As can be seen in this figure, close encounters between the secondary star (body B) and the planetary object (body C) greatly increase the numerical error. For encounter distances below $\sim$20 AU, the conservation of the Jacobi Constant is severely degraded. On the other hand, close approaches between the primary (body A) and the planetary body do not seem to markedly degrade Jacobi Constant conservation. This is presumably because body C is massless. Consequently, the democratic heliocentric coordinates are equivalent to heliocentric coordinates, and $H_{\rm jump}$ is zero.

\begin{figure}
\centering
\includegraphics[scale=0.5]{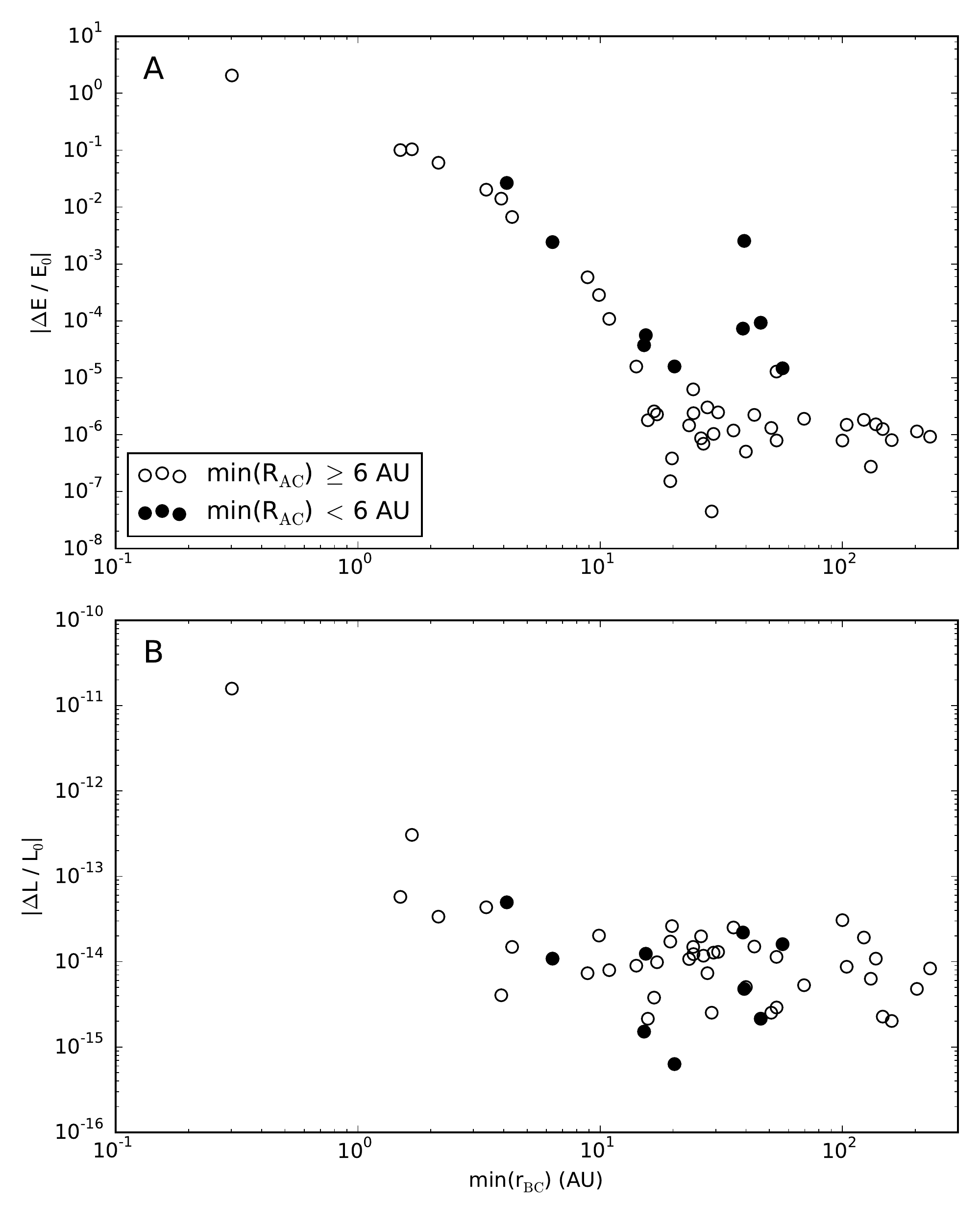}
\caption{{\bf A}: Fractional change in energy between the beginning and end of 50 simulations of a 1.92 M$_{\sun}$ star (body A), a 0.73 M$_{\sun}$ star (body B), and a 1 M$_{Jup}$ planetary object (body C). This is plotted against the minimum approach distance of bodies B and C. Data points from simulations in which bodies A and C have an approach within 6 AU are filled. {\bf B}: Fractional change in angular momentum for the same set of simulations.  Data points from simulations in which bodies A and C have an approach within 6 AU are again filled. }
\label{fig:starplELerr}
\end{figure}

As with are our study of the accuracy of stellar dynamics, we now repeat our set of 50 simulations one more time, again until the planetary body (C) is ejected. In these reruns, the orbits of A, B, and C are the same, but now body C is given a non-zero mass of 1 M$_{Jup}$. In our reruns, we measure the conservation of energy and angular momentum. The results of these simulations are shown in Figure \ref{fig:starplELerr}. Just as with our Jacobi Constant analysis, we see that close encounters between B and C within $\sim$20 AU significantly degrade the conservation of energy. However, we now also see that encounters between A and C also increase numerical error. From Figure \ref{fig:starplELerr}A, it appears that energy conservation becomes substantially worse if C passes within $\sim$6 AU of A. Finally in Figure \ref{fig:starplELerr}B we see that very close encounters between the secondary star and planetary body can degrade the conservation of angular momentum, but the fractional error still remains below $10^{-10}$ for all of our simulations.

To this point, we have characterized the accuracy of our code using unstable systems that dissolve on $\sim$Myr timescales. This has enabled us to quantify the error resulting from close encounters between bodies of various classes over short timescales, but we have not studied how our code behaves modeling stable systems over long timescales. To do this, we assemble 50 systems of well-spaced spaced bodies with a clear hierarchy. For our stellar components, we again mimic the Fomalhaut system masses by having a 1.92 M$_{\sun}$ primary, a 0.73 M$_{\sun}$ secondary, and a 0.18 M$_{\sun}$ tertiary star. The secondary stars of our systems are placed on orbits with semimajor axes of 2,000 AU, eccentricities drawn randomly between 0 and 0.1, and inclinations drawn randomly between 0 and 10$^{\circ}$. The tertiary stars of our systems are assigned similarly drawn eccentricities and inclinations (about the center-of-mass of the primary and secondary) and given semimajor axes of 20,000 AU. In addition to the stellar bodies of this system, two planetary bodies are placed in orbit about the primary star. The first is a Jupiter-mass planet on a circular orbit at 50 AU, and the second is a Saturn-mass planet on a circular orbit 10 mutual Hill radii further from the primary at 92.8 AU. Both planets are given inclinations between 0 and 1$^{\circ}$. Our $T+V$ scheme within inertial coordinates is used to integrate the secondary and tertiary stars, while a mixed variable symplectic routine in democratic heliocentric coordinates is used for the planetary integrations.

\begin{figure}
\centering
\includegraphics[scale=0.5]{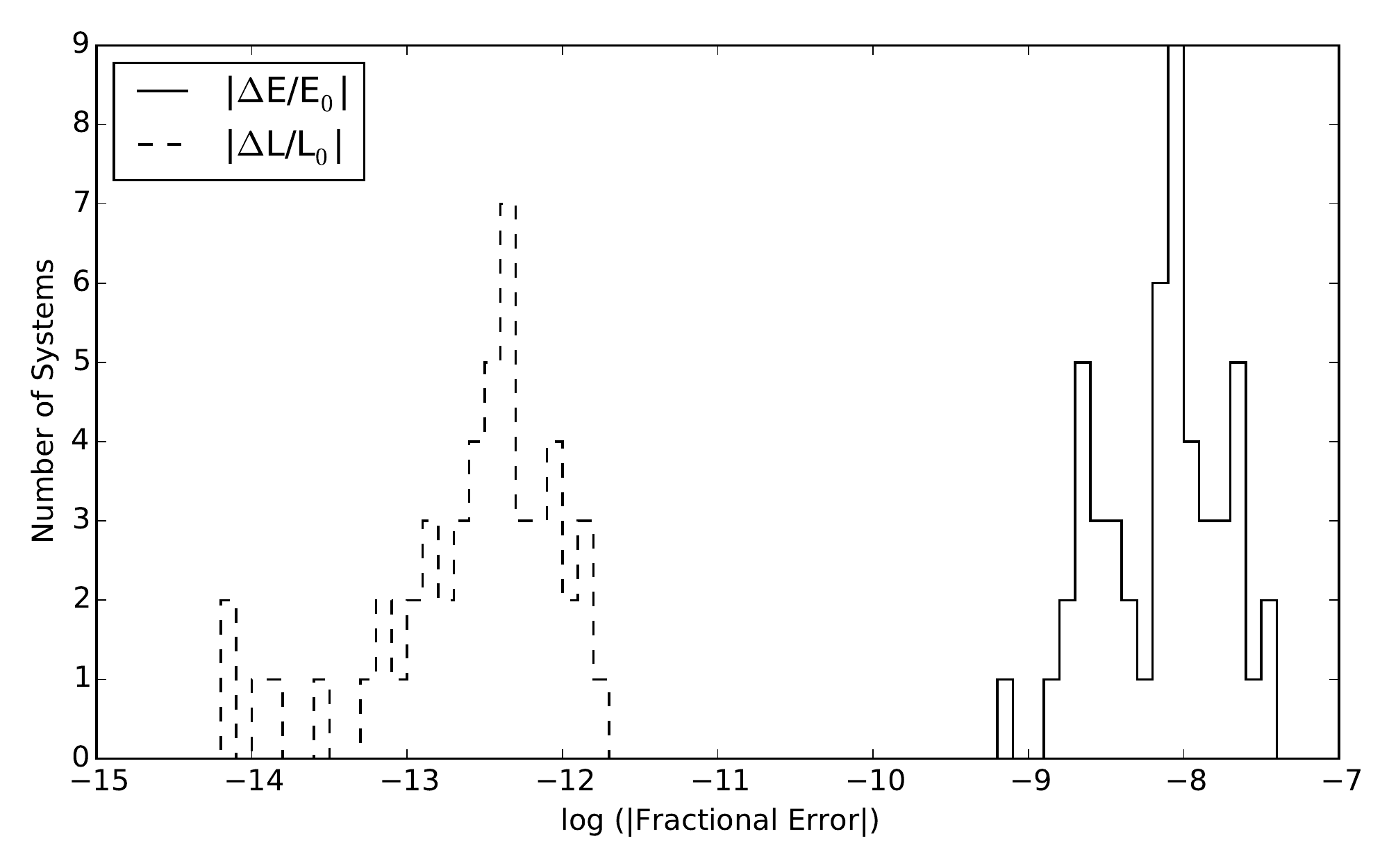}
\caption{Distribution of the fractional change in energy {\it solid} and angular momentum {\it dashed} over the course of 500-Myr integrations of 50 stable systems consisting of a 1.92 M$_{\sun}$ primary star orbited by a Jupiter-mass planet, a Saturn-mass planet, a distant 0.73 M$_{\sun}$ stellar companion, and an even more distant 0.18 M$_{\sun}$ stellar companion.}
\label{fig:stableELerr}
\end{figure}

Our 50 systems are integrated for 500 Myrs, and the final energy and angular momentum of each system is compared with its initial values. The absolute fractional changes in energy and angular momentum are shown in Figure \ref{fig:stableELerr}. As can be seen, both quantities are conserved to a very high degree in all of our systems. We have no systems where energy conservation is worse than 1 part in 10$^{7}$ and none where angular momentum conservation is worse than 1 part in 10$^{11}$. 

Based on Figures \ref{fig:secondaryerr}--\ref{fig:stableELerr}, we can conclude that the only significant source of error in our Fomalhaut simulations will be when stellar companions pass within $\sim$20 AU of the primary, when stellar companions pass within $\sim$20 AU of a planetary body orbiting the primary, or when planetary bodies pass within $\sim$6 AU of the primary. This enables our simulations to accurately determine whether Fomalhaut A's stellar companions could have distorted the shape of its debris belt. Because the ring orbits at over 100 AU, if belt particles attain pericenters below 6 AU, they will have eccentricities of well over 0.9. Meanwhile, the observed eccentricity of Fomalhaut's belt is $\sim$0.1, so such highly eccentric ring particles will not be considered matches to the system anyways. 

One final way to quantify our code's accuracy that is very relevant to our work here is to simulate the orbital excitation of a belt of planetesimals as they are perturbed by stellar passages. To do this, we run a set of simulations modeling the evolution of 100 belt particles as they are perturbed by a single stellar passage. The belt particles consist of massless bodies placed on near circular ($e<10^{-3}$), coplanar ($i<1^{\circ}$) orbits between 127 and 143 AU around a primary with a mass of 1.92 M$_{\sun}$. With this architecture in place, the system is subjected to a single flyby of a 0.73 M$_{\sun}$ star on a parabolic orbit with a random spatial orientation. Fifty different simulations of single flybys are preformed at flyby distances of 100, 125, 150, 175, 200, 250, 300, 400, 500, 600, 700, 800, 900, 1000, 1200, 1400, 1600, 1800, and 2000 AU. 

\begin{figure}\centering
\includegraphics[scale=0.5]{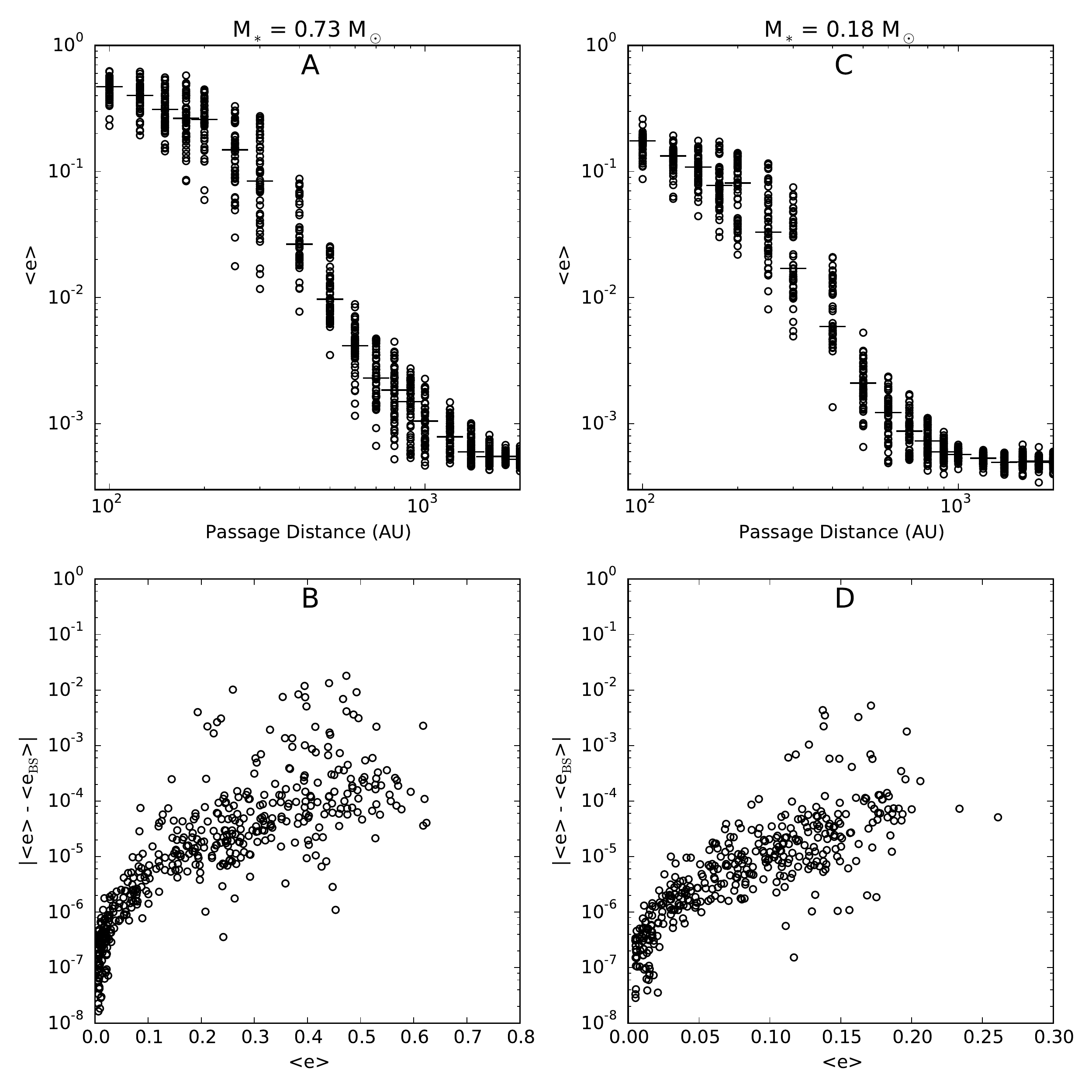}
\caption{{\bf A:} Ring particles on initially circular orbits are subjected to passages of a 0.73 M$_{\sun}$ star. The final median orbital eccentricity of ring particles is plotted against the distance of the stellar passage to which they are subjected. Circular data points mark the results of individual stellar passages and horizontal lines mark the ``median of median values'' for each stellar passage distance. {\bf B:} Absolute difference between the median ring eccentricity predicted by our symplectic integrations from Panel A and that predicted by a Bulirsch-Stoer integration of the same stellar passage is plotted against the symplectic median ring eccentricity for each one of our simulations. {\bf C--D:} Analogous plots to Panels A--B, but for stellar passages involving a 0.18 M$_{\sun}$ star. }
\label{fig:starpasserr}
\end{figure}

In Figure \ref{fig:starpasserr}A, we plot the final median eccentricity of the belt in each simulation as a function of the stellar passage distance. Here we see that even for stellar passages at 200 AU, most flybys result in the belt being perturbed to a more eccentric state than the observed one. Because the radial extent of the belt is $\sim$50 AU closer to the primary than the stellar passage distance, these types of encounters exclude the possibility of significant numerical error due to a planetary body passing within 20 AU of the 0.73 M$_{\sun}$ star. Thus, before numerical errors become a possible issue in our simulations, the belt will have already been excited to an eccentricity comparable to or exceeding the observed value. In our study of the Fomalhaut system, we are ultimately interested in the overall probability that the stellar companions can excite the eccentricity of the belt. Whether this excitation yield a belt eccentricity of 0.1 or 0.35 is not as critical. 

To verify the accuracy of the stellar passage simulations in Figure \ref{fig:starpasserr}A, we repeat them one more time with a Bulirsch-Stoer integrator. In Figure \ref{fig:starpasserr}B, we compare the median belt eccentricities found in the Bulirsch-Stoer runs with the eccentricities found in the runs using our symplectic code. We see that the difference in median eccentricity is well below 10$^{-3}$ for any stellar flyby capable of exciting the belt to a state comparable to its observed eccentricity. Even at flybys that yield eccentricities exceeding 0.5, the eccentricity error is still only $\sim$0.01--0.02 at most. The simulations with these largest errors seem to correspond to stellar passages at or below 200 AU, where the close encounters between planetary bodies and stellar companions that yielded large errors in Figures \ref{fig:starplerr} and \ref{fig:starplELerr} become possible. Even in these cases, 99\% of our simulations have eccentricity differences below 0.01 when comparing the two different runs. Similarly, if we only select stellar passages that yield eccentricities above the observed ring value ($e > 0.15$) we again find that the eccentricity difference in our two sets of runs is below 0.01 in 99\% of our systems. Thus, when our simulations predict that Fomalhaut B significantly excites the belt of Fomalhaut A, we can trust that this is indeed the case.

In our simulations of the Fomalhaut system, it is also possible that Fomalhaut C is the star that excites the eccentricity of the Fomalhaut A belt. To gauge the types of encounters necessary for this, we again perform 50 different simulations of single stellar flybys at distances of 100, 125, 150, 175, 200, 250, 300, 400, 500, 600, 700, 800, 900, 1000, 1200, 1400, 1600, 1800, and 2000 AU. However, this time the stellar mass used is that of Fomalhaut C, or 0.18 M$_{\sun}$. The results of these simulations are shown in Figure \ref{fig:starpasserr}C, where we see that single encounters within at least 300 AU are necessary to perturb the belt to its observed state. Even many encounters at 100 AU do not excite the belt beyond its observed eccentricity.

Once again, we also repeat our stellar passage simulations with a Bulirsch-Stoer integrator and examine the difference in predicted median belt particle eccentricities for our two different integrations in Figure \ref{fig:starpasserr}D. As with our other stellar passage experiments, we find that the difference in predicted median belt eccentricity is typically small. In this case, the difference is always below 0.005. If we limit ourselves to simulations with median eccentricities above 0.15, we find that the difference in predicted eccentricity is below 0.001 in 94\% of our pairs of integrations. As with Fomalhaut B, our simulations should do an excellent job of flagging when Fomalhaut C does and does not excite the eccentricity of the belt around Fomalhaut A. 

\clearpage

\bibliographystyle{apj}
\bibliography{Fomalhaut}

\end{document}